\documentclass[10pt,journal,compsoc]{IEEEtran}

\usepackage{amsmath,amssymb,amsfonts, amsthm}
\usepackage{algorithmic}
\usepackage{graphicx}
\usepackage{textcomp}

\usepackage[backend=biber, style=ieee]{biblatex}
\usepackage{array}
\usepackage[caption=true,font=footnotesize,labelfont=sf,textfont=sf]{subfig}
\usepackage{tcolorbox}
\usepackage{url}
\usepackage[hidelinks]{hyperref}
\usepackage{wrapfig}
\theoremstyle{definition}
\newtheorem{definition}{Definition}

\usepackage{orcidlink}
\usepackage[noabbrev,nameinlink]{cleveref}
\bibliography{bibliography}  
\usepackage{oz}
\renewcommand{\itemautorefname}{\@gobble}
\usepackage{xspace}
\usepackage{eurosym}
\usepackage{stfloats}

\newcommand{\dt}{digital twin\xspace}
\newcommand{\dts}{digital twins\xspace}
\newcommand{\pt}{physical twin\xspace}
\newcommand{\pts}{physical twins\xspace}
\newcommand{\ifz}{Industry 4.0\xspace}
\newcommand{\dtp}{digital twin prototype\xspace}
\newcommand{\dtps}{digital twin prototypes\xspace}

\newcommand{\pubsub}{publish/subscribe\xspace}
\newcommand{\hil}{HIL\xspace}
\newcommand{\sil}{SIL\xspace}

\begin{document}

\title{From Digital Twins to Digital Twin Prototypes: Concepts, Formalization, and Applications}

\author{
\IEEEauthorblockN{Alexander Barbie\IEEEauthorrefmark{1} and Wilhelm Hasselbring\IEEEauthorrefmark{1}} \\ 
\vspace{1em}
\IEEEauthorblockA{\IEEEauthorrefmark{1}Software Engineering Group, Christian-Albrechts-University, Kiel (Germany)}
}


\IEEEtitleabstractindextext{%
\begin{abstract}
The transformation to Industry 4.0 also transforms the processes of how we develop intelligent manufacturing production systems. To advance the software development of these new (embedded) software systems, digital twins may be employed. However, there is no consensual definition of what a digital twin is. In this paper, we give an overview of the current state of the digital twin concept and formalize the digital twin concept using the Object-Z notation. This formalization includes the concepts of physical twins, digital models, digital templates, digital threads, digital shadows, digital twins, and digital twin prototypes. The relationships between all these concepts are visualized as UML class diagrams.

Our digital twin prototype (DTP) approach supports engineers during the development and automated testing of complex embedded software systems. This approach enable engineers to test embedded software systems in a virtual context, without the need of a connection to a physical object. In continuous integration / continuous deployment pipelines such digital twin prototypes can be used for automated integration testing and, thus, allow for an agile verification and validation process.

In this paper, we demonstrate and report on how to apply and implement a digital twin by the example of two real-world field studies (ocean observation systems and smart farming).
For independent replication and extension of our approach by other researchers, we provide a lab study published open source on GitHub.
\end{abstract}

}

\maketitle

\IEEEdisplaynontitleabstractindextext

\section{Introduction}\label{sec:introduction}
For cyber-physical-systems, the Industrial Internet of Things (IIOT), and Industry 4.0 applications, the embedded software is an increasingly crucial asset. With increasing requirements and hence, increasing complexity, new challenges arise for manufacturers and in particular, for the engineers of these systems. While in large software companies, software development is often done by distributed teams of engineers \cite{Jackson2022}, this is usually different for small and medium-sized enterprises (SME) that develop embedded systems~\cite{interviewpaper}. Especially, in SMEs, embedded software still is often developed by the same engineers who also develop the electronics and/or mechanical parts~\cite{ICSA2019}. 

However, with the demand for context-aware, autonomous, and adaptive robotic systems \cite{DTroadmap}, more advanced software engineering methods have to be adopted by the embedded software community. Consequently, the way these systems are developed has to advance. In future development workflows, the embedded software systems will be the center-piece of IIoT applications. To achieve this, the community has to move from expert-centric tools \cite{DTroadmap} to modular systems, whereby domain experts are enabled to contribute parts of the system.

A survey among 2,000 decision makers about trends and challenges in software engineering found that \emph{quality} is perceived in the software industry as the single most relevant premise to survive \cite{50yearsse}. Yet, organizations struggle to achieve software quality along with cost and efficiency \cite{SIsoftwarequality}. 
During the development of embedded (software) systems, at some point, thorough and reliable tests are necessary to verify and validate the whole system \cite{DTElectricalVehicles}. A common way to test the control algorithms of an embedded software system is Hardware-in-the-Loop (HIL) testing. An example for HIL testing at large scale is Airbus with creating iron birds of their aircraft, containing the corresponding electronics, hydraulics and flight controls \cite{ironbird}. However, many SMEs cannot afford such redundant hardware just for the purpose of testing software. Hence, test \emph{automation} is among the most popular topics for testing embedded software \cite{studyembeddedtesting}.
Still, automatic quality assurance is a challenge in this context, since hardware is in the loop.

Many different simulation tools were proposed, developed, and sold, with the promise to reduce costs and time needed for verification and validation. Yet, none of these tools is able to combine all aspects of modern machines during all steps of the production life-cycle, due to the complexity of systems and the high amount of data being processed. Thus, multidisciplinary simulation concepts are increasingly important with regard to scalable and highly modular production environments enabled by cyber-physical systems~\cite{cpssimulationmanufacturing}. Alongside HIL testing, manufactures implemented different automated testing strategies with In-the-Loop simulations to reduce costs, e.g., Software-in-the-Loop (SIL), Model-in-the-Loop (MIL), and Processor-in-the-Loop (PIL) simulations \cite{Bringmann2008}.

One promising technique to enhance the overall software quality of embedded systems, is the Digital Twin concept. We start with a discussion of related work in \Cref{sec:relatedwork}. As there is no common understanding around the concept, we then dissect the different parts of a digital twin in and formally specify the concepts with the Object-Z notation. Afterwards, the application of digital twins in different industrial contexts are presented to illustrate the approach.

\section{Related Work}\label{sec:relatedwork}
Digital twins are not only a growing topic in academia but also in the industry, especially in manufacturing \cite{SaraccoSIIntroduction}. However, there is still no consensual definition of a Digital Twin, as we explain in \Cref{subsec:dtevolution}. Most of the research conducted to find a general definition of a \dt, are literature reviews \cite{Negri2017, dtdef-tao, dtdef-rosen, dtdef-kritzinger} investigating where \dts are used, which components are part of it, and which level of integration with the CPS exists. In particular, \textcite{dtdef-kritzinger} contributed with their literature review to a consensual understanding about which subsystems are part of a \dt. They consider the digital model, the digital shadow, and the \dt as three separate levels of integration in the overall concept of \dts. In this paper, we extend this work by providing a formalization for all these categories.

With regard to mathematical approaches to formalize the concept of \dts, there is a lack in research papers. Nevertheless, we discuss two approaches \cite{Yue2021, Becker2021} that use semi-formal approaches to define the relationships between the different components of a \dt in \Cref{subsec:semiformal}.

\subsection{The Evolution of the Digital Twin Concept}\label{subsec:dtevolution}
An innovative method for testing and monitoring embedded systems was used for space missions, dating back to the early Apollo missions conducted by the National Aeronautics and Space Administration (NASA). Here, the ``Twin'' concept was initially employed during the Apollo missions in the late 1960s as a safety precaution. If a system on the spacecraft failed during the mission, engineers had no access to the capsule. A failure to fix problems in a timely manner could be catastrophic for the space mission. At the time, computational power was insufficient for complex simulations, so NASA engineers came up with the idea of building at least two identical space capsules. One was used for the mission while the other remained on Earth, serving as the ``Twin'' for simulation purposes. Changes to the system were first tested on the Twin before astronauts received instructions. This approach required both capsules to be maintained exactly the same, including replacing parts on the Twin even if it was not used during a mission. NASA had planned to transfer this approach to the Space Shuttle program, but abandoned the idea due to the high costs.

\begin{figure}[ht]
    \centering
    \includegraphics[width=.45\textwidth]{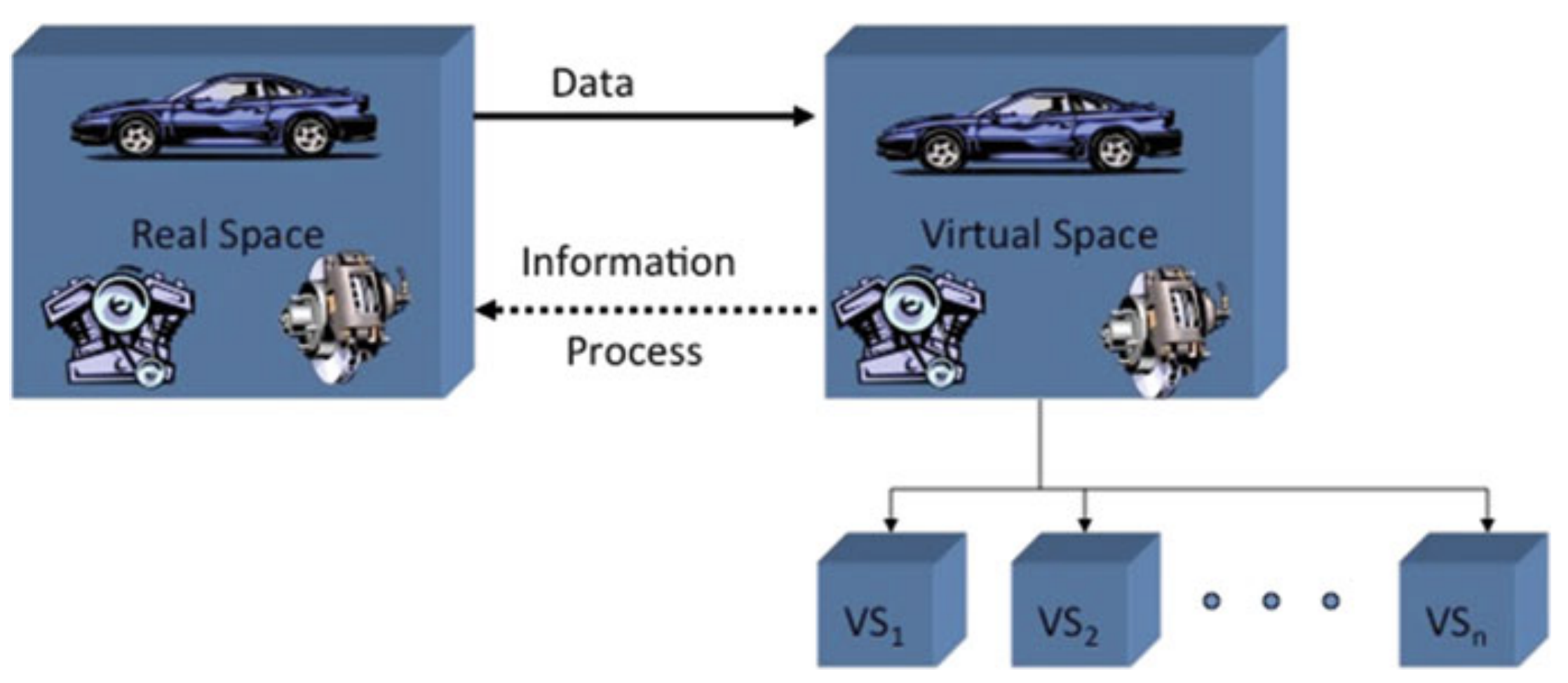}
    \caption{A Digital Twin by \textcite{dtdef-grieves} consists of the real space (left side), the virtual space (right side), and the link for data flow from real space to virtual space. The opposite direction is done manually by using information to enhance processes (Source: \cite{dtdef-grieves}).}
    \label{fig:grievesdt}
\end{figure}

Half a century later, with advancements in computational power and improved simulations, the NASA's Twin concept has evolved into a \dt. However, there was a second research threads that contributed to the concept. The second thread originated from the manufacturing industry and dates back to 2002, when \textcite{grievesorigin} first pitched for the formation of a Product Lifecycle Management (PLM) center at the University of Michigan. The presentation slide, as depicted in \Cref{fig:grievesdt}, had the title ``Conceptual Ideal for PLM'' \cite{dtdef-grieves} and sketched the idea of a \dt and named it ``Mirrored Spaces Model'' back than \cite{grievesorigin}.

Grieves envisioned with the Mirrored Spaces Model already three crucial components of  \dts: the physical space, the virtual space, and the data link between the physical and virtual spaces. 
Later, in 2016, \textcite{dtdef-grieves} defined the \dt as stated in \Cref{def:dtdef-grieves}: 
\begin{tcolorbox}[colback = white]
\begin{definition}[Digital twin by \textcite{dtdef-grieves} (2016)]\label{def:dtdef-grieves}
The Digital Twin is a set of virtual information constructs that fully describes a potential or actual physical manufactured product from the micro atomic level to the macro geometrical level. At its optimum, any information that could be obtained from inspecting a physical manufactured product can be obtained from its Digital Twin. Digital Twins are of two types: Digital Twin Prototype (DTP) and Digital Twin Instance (DTI). Digital twin’s are operated on in a Digital Twin Environment (DTE).        
\end{definition}
\end{tcolorbox}
\Cref{def:dtdef-grieves} considered the \dt to be a collection of technologies and distinguished between two types: the Digital Twin Prototype (DTP) and the Digital Twin Instance (DTI). The Digital Twin Prototype is a set of blueprints, etc., used to construct or maintain the \pt. The Digital Twin Instance is the specific instance created after the \pt has been manufactured and is linked to it throughout its lifecycle.
Although the vision by \textcite{dtdef-grieves} reflected solutions that are possible today, the technology available in 2002 only allowed for a rudimentary implementation of what a \dt is known today. Digital twins were seen as a new paradigm for designing, manufacturing, and servicing products \cite{SaraccoSIIntroduction}. However, the meaning of \dt may vary depending on the sector they are utilized in \cite{SaraccoSIIntroduction}.

After their introduction, \dts experienced a hype phase until around the year 2006. The first hype of \dts was driven by high hopes in the industry. However, the technology did not live up to the hype, and \dts became a buzzword in marketing departments rather than a fully realized concept. \textcite{newman_building_2021} observed and criticized something similar with regard to microservice architectures. \textcite{SaraccoSIIntroduction} emphasize that the industry drove the development of \dts, while academia ignored it. The revival of interest in \dts in 2016 was thanks to the maturity of IIoT and CPS technologies, and academia also joined the bandwagon. Digital twins reached the peak of the Gartner Hype Cycle of emerging technologies in 2018 \cite{gartnerhypecycle2018dt}. Furthermore, an increased number in research papers and special issues published by journals can be registered after 2016.

It was between 2006 and 2016 when \textcite{nasa-first}, and \textcite{dtdef-nasa} proposed their vision for a \dt for NASA \cite{grievesorigin}. \textcite{nasa-first} used the term \dt in their technology roadmap for NASA. However, they described the \dt concept, but did not define \dts. The better known \dt definition was by \textcite{dtdef-nasa} for next generation fighter aircraft and NASA vehicles shown in \Cref{def:dtdef-nasa}:
\begin{tcolorbox}[colback = white]
    \begin{definition}[Digital twin by \textcite{dtdef-nasa} (NASA) (2012)]\label{def:dtdef-nasa}
A Digital Twin is an integrated multiphysics, multiscale, probabilistic simulation of an as-built vehicle or system that uses the best available physical models, sensor updates, fleet history, etc., to mirror the life of its corresponding flying twin. The Digital Twin is ultra-realistic and may consider one or more important and interdependent vehicle systems, including airframe, propulsion and energy storage, life support, avionics, thermal protection, etc. 
\end{definition}
\end{tcolorbox}
They tailored their vision for the specific use case of spacecraft, satellites, and space exploration, where simulations play a crucial role due to the high cost of hardware and human resources. These simulations are used both in the development phase, which indicates at least a MiL approach, and to monitor the systems during missions. To detect anomalies during flight, they also included a channel for sending sensor data from the \pts to their corresponding \dts. Loading this data into the simulation with a realistic model supersedes the NASA's Twin approach from the Apollo missions. This is similar to the data link shown in \Cref{fig:grievesdt}, only with far advanced technology and tools. A demonstration of their implementation can be seen in the Perseverance Rover that landed on Mars in 2021 \cite{perseverancemultimedia}.

In parallel to the definition by NASA, \textcite{dtdef-garetti} defined \dts for manufacturing as shown in \Cref{def:dtdef-garetti}:
\begin{tcolorbox}[colback = white]
\begin{definition}[Digital twin by NASA \cite{dtdef-garetti} (2012)]\label{def:dtdef-garetti}
The \dt consists of a virtual representation of a production system that is able to run on different simulation disciplines that is characterized by the synchronization between the virtual and real system, thanks to sensed data and connected smart devices, mathematical models and real time data elaboration. The topical role within \ifz manufacturing systems is to exploit these features to forecast and optimize the behaviour of the production system at each life cycle phase in real time.
\end{definition}
\end{tcolorbox}

When the attention on \dts research rekindled, academia proposed multiple definitions for the concept \cite{Negri2017}. These definitions were influenced by the realistic simulation approach put forth by NASA. \textcite{dtdef-rosen} linked the \dt concept to the \ifz strategy of the German Platform Industry 4.0~\cite{plattform40}. They illustrated how simulations evolved over time, from mechanics in the 1960s to simulation-based system design and finally to \dts since 2015. They also highlighted that modularity, autonomy, and connectivity are crucial requirements for \dts, among other factors.

The definitions provided by \textcite{dtdef-grieves} and NASA only included an automated connection from the \pt to its \dt. \textcite{dtdef-trauer} conducted an industrial case study to analyze how the industry perceived and defined \dts between 2002 and 2019. They traced the evolution of \dts and presented \Cref{def:dtdef-trauer} as a result.

\begin{tcolorbox}[colback = white]
    \begin{definition}[Digital twin by \textcite{dtdef-trauer} (2020)]\label{def:dtdef-trauer}
A Digital Twin is a virtual dynamic representation of a physical system, which is connected to it over the entire life cycle for bidirectional data exchange.
    \end{definition}
\end{tcolorbox}

\begin{figure*}[ht]
\centering
    \subfloat[\label{subfig:digitalmodel}]{%
      \includegraphics[width=0.3\textwidth]{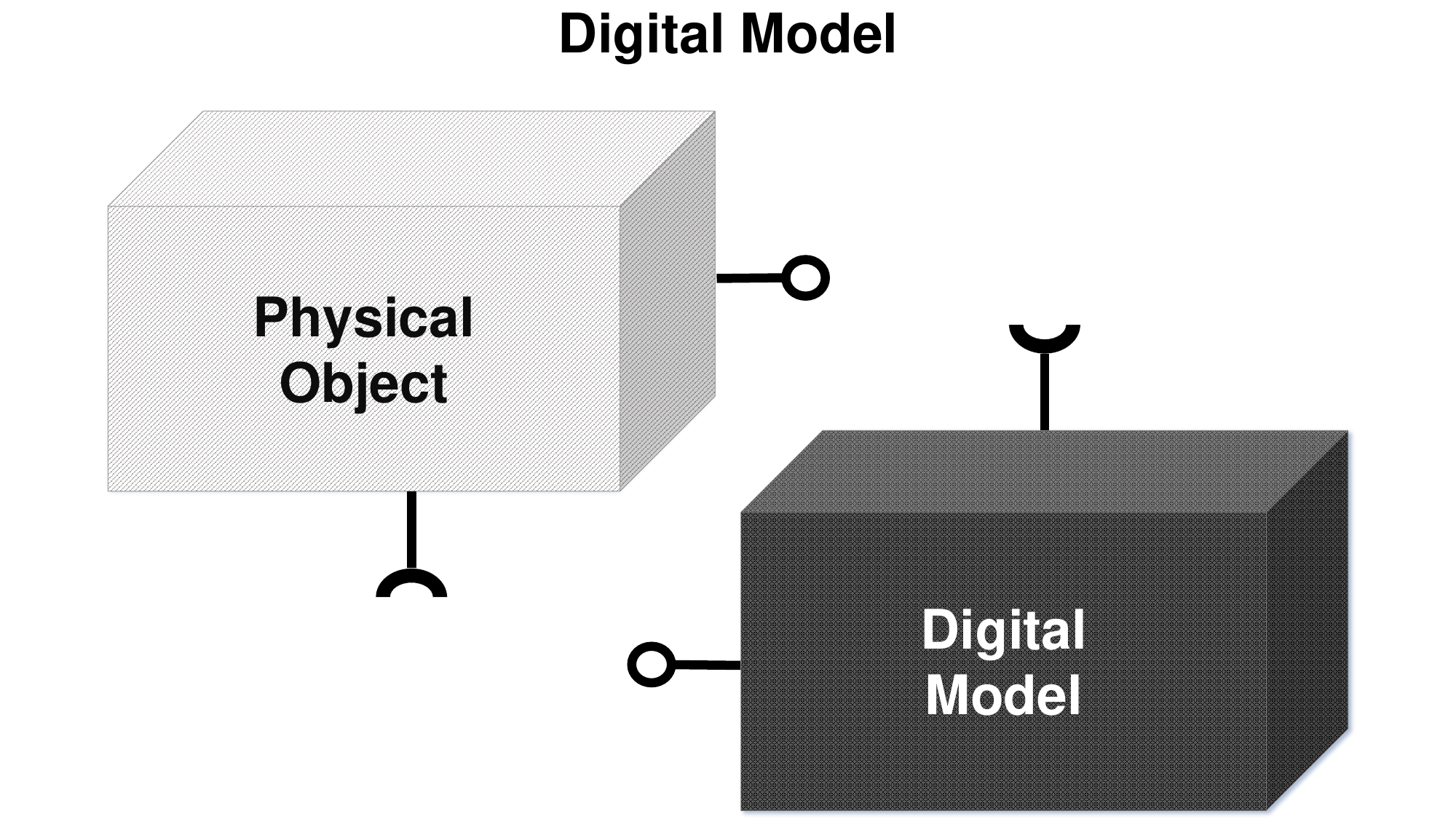} 
    }
    \hfill
    \subfloat[\label{subfig:digitalshadow}]{%
      \includegraphics[width=0.3\textwidth]{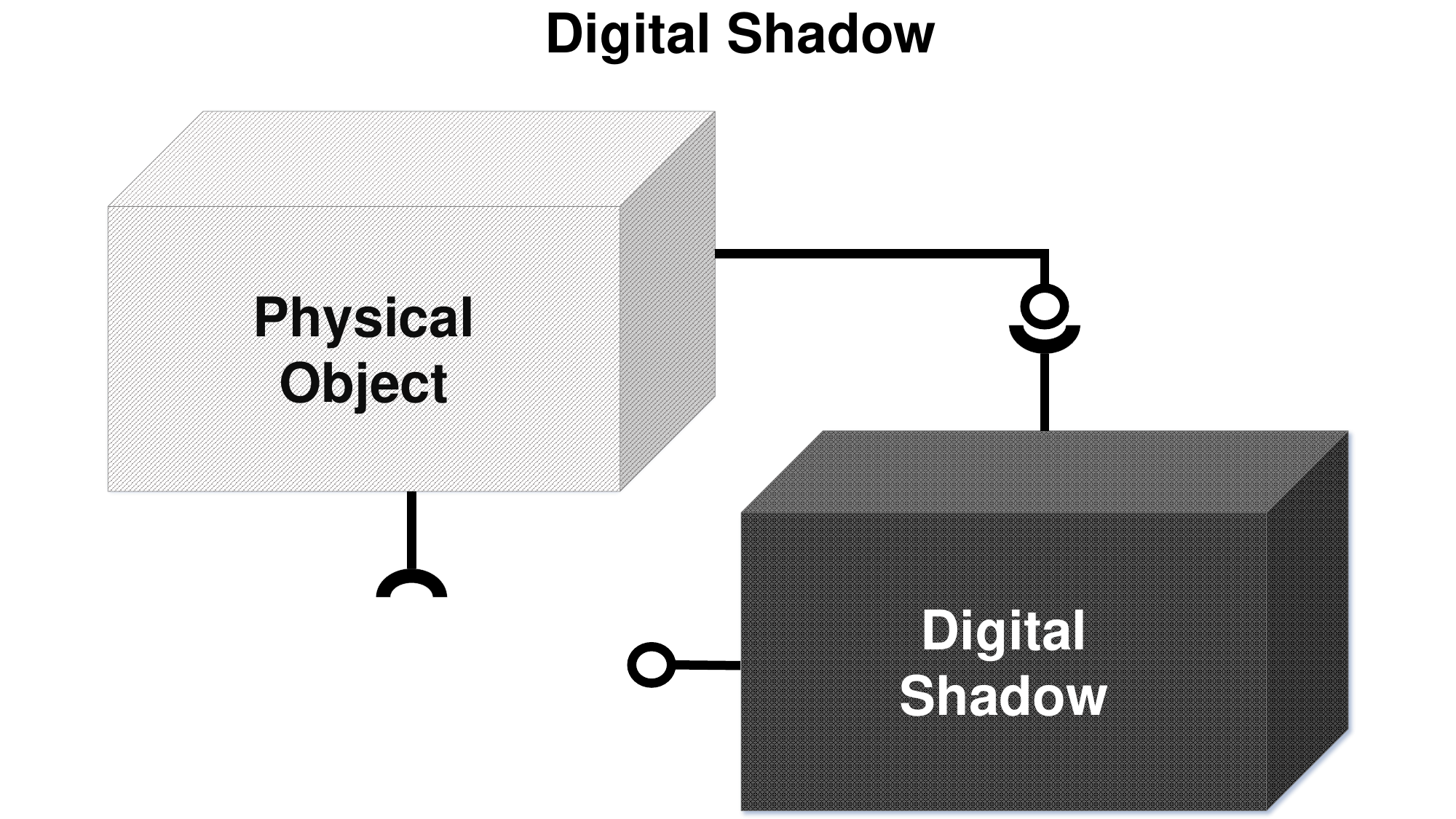} 
    }
    \hfill
    \subfloat[\label{subfig:digitaltwin}]{%
      \includegraphics[width=0.3\textwidth]{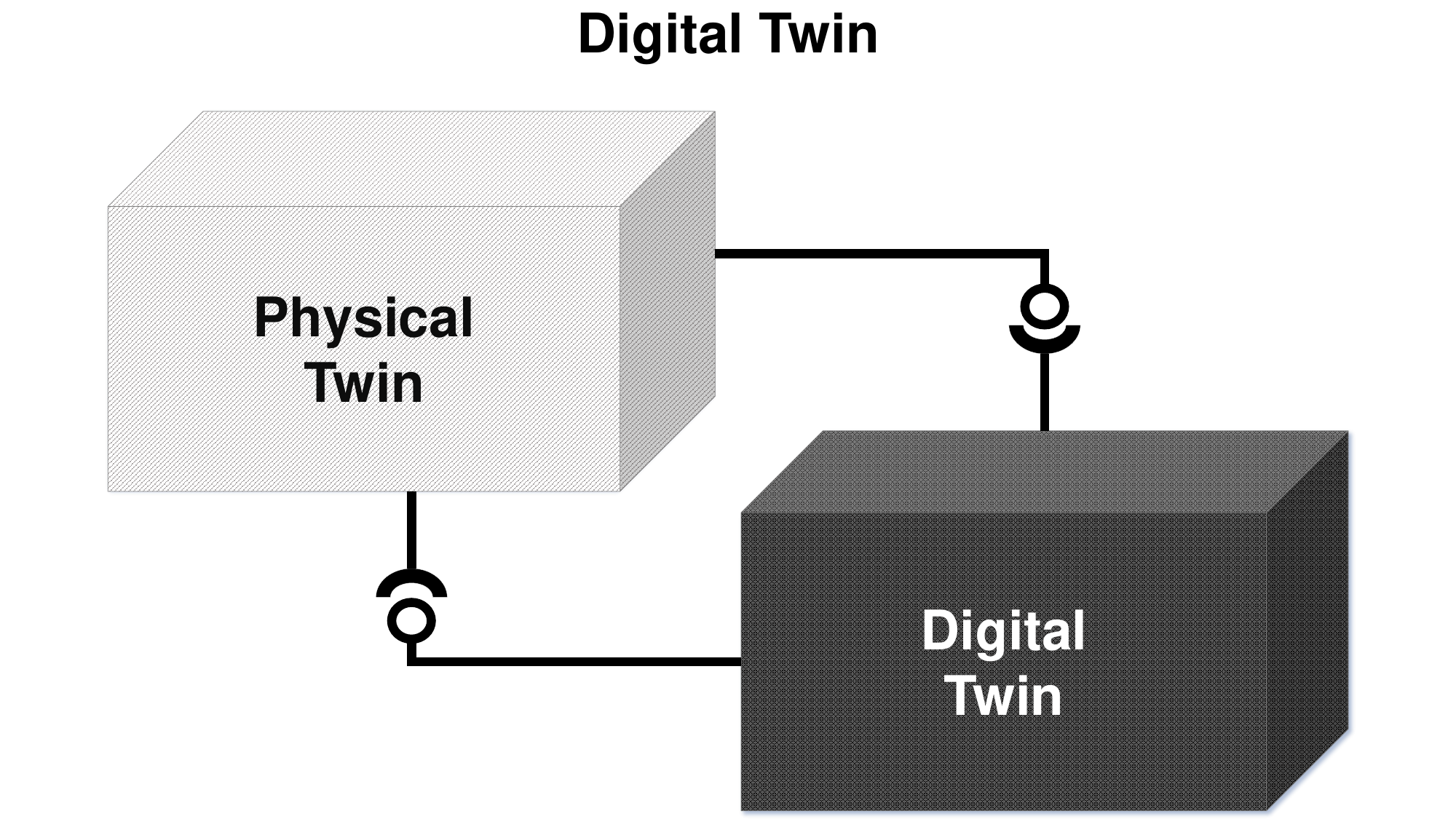}
    }
    \caption{Subcategories of digital twins by their level of integration with the \pts (Source: \cite{dtdef-kritzinger}).}
    \label{fig:dtboxs}
\end{figure*}

We present \Cref{def:dtdef-trauer} here, because of the inclusion of the \emph{bidirectional} data exchange from \dt to \pt. This \emph{bidirectional} interaction allows remote control and operation of the \pt, as well as new opportunities for collaboration between \pt and \dt. This poses a challenge for engineers to either develop the software independently for each twin, violating the principle of realistic replication, or to use tools like Docker to containerize the \pt's software for use as a \dt.

Depending on the research field, the industry, and use cases, the term \dt is often used synonymous with concepts like \emph{Digital Model}, \emph{Digital Shadow}, and \emph{Digital Thread} \cite{Negri2017, dtdef-kritzinger}. \textcite{dtdef-kritzinger} conducted a categorical literature review and analyzed research papers with regard of the proposed concept and how it deviates from a common understanding of the essential parts of \dts. They classify three subcategories of a \dt by their level of integration with the \pt: (i) digital model, (ii) digital shadow, and (iii) \dt. The differences are depicted in \Cref{fig:dtboxs}.

\begin{itemize}
    \item \Cref{subfig:digitalmodel} shows the \emph{digital model}. There is no automated connection between the physical object and the digital model. No automated data exchange is realized. State changes in the physical object do not immediately affect the digital model and vice versa.
    \item If there is an automated one-way data flow from the physical object to the digital object (see \Cref{subfig:digitalshadow}), then this is a \emph{digital shadow}. A change in state of the physical object leads to a change of state in the digital shadow, but not vice versa. 
    \item \Cref{subfig:digitaltwin} shows a fully integrated \emph{digital twin}. The data flows are automated between the \pt and the digital twin in both directions. In such a configuration, the digital twin might also act as a controlling instance of the \pt. A change in state of the \pt directly leads to a change in state of the digital twin and vice versa.
\end{itemize}

With the increasing importance of \dts, the International Organization for Standardization (ISO) also published the ISO~23247 series, defining a framework to support the creation of \dts of observable manufacturing elements, including personnel, equipment, materials, manufacturing processes, facilities, environment, products, and supporting documents \cite{dtdef-iso23247}.

\begin{tcolorbox}[colback = white]
    \begin{definition}[Digital twin by \textcite{dtdef-iso23247} (2021)]\label{def:dtdef-iso23247}
A digital twin assists with detecting anomalies in manufacturing processes to achieve functional objectives such as real-time control, predictive maintenance, in-process adaptation, Big Data analytics, and machine learning. A digital twin monitors its observable manufacturing element by constantly updating relevant operational and environmental data. The visibility into process and execution enabled by a digital twin enhances manufacturing operation and business cooperation   
    \end{definition}
\end{tcolorbox}
One aspect of ISO~23247 that immediately catches the eye is the absence of mentioning of bidirectional communication. The focus is on the monitoring aspect of a \dt. According to the definition by \textcite{dtdef-kritzinger}, ISO~23247 only describes a digital shadow \cite{dtdef-iso23247}.

\begin{figure*}[b]
    \centering
    \includegraphics[width=.6\textwidth]{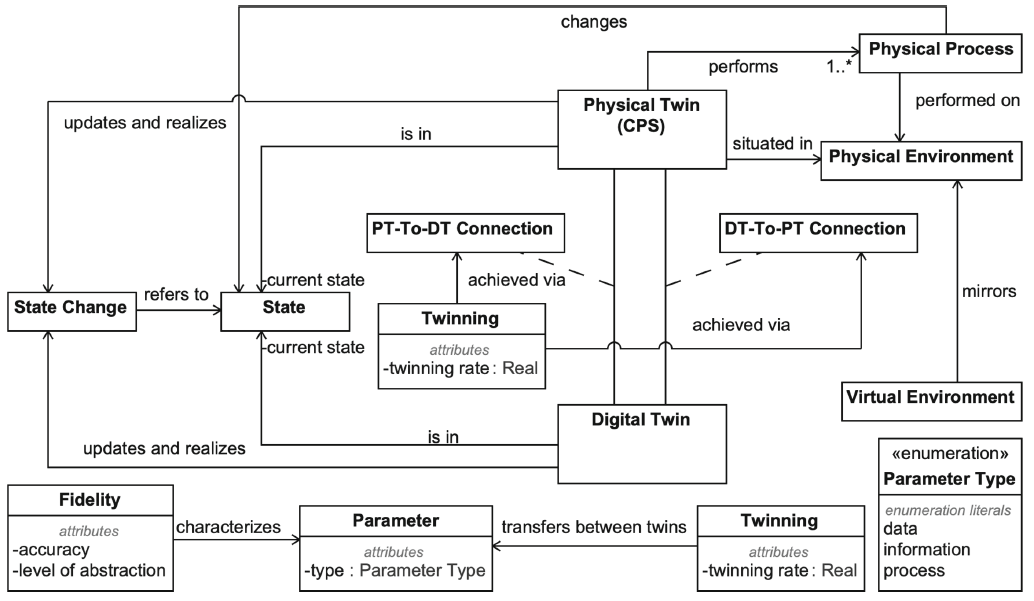}
    \caption{Semi-formal description of the relationships between physical twin, digital twin, their connections, and environments as described by \textcite{Yue2021}.}
    \label{fig:awsuml}
\end{figure*}

Since 2018, IIoT platforms transitioned from basic data hubs to digital twin (DT) platforms. \textcite{Lehner2022} evaluated the \dt platforms provided by Amazon Web Services (AWS), Microsoft Azure, and the Eclipse ecosystem and showed that they fulfill many requirements, yet not all key requirements. Features like bidirectional synchronization between physical and digital twins require additional coding, and automation protocols are not covered yet. According to the categorization of the integration level of \dts \cite{dtdef-kritzinger}, these platforms only help to establish a so-called \emph{digital shadow}~\cite{dtdef-kritzinger}. Modern simulation tools such as AutoDesk, aPriori, or Ansys, are using IIoT platforms to feed the simulation with data and enable the integration of automation protocols. Often they are promoted with the promise of a digital twin. However, similar to the cloud providers, these tools also just help to establish a digital shadow. The simulation of a \pt (PT) still does not cover the entire embedded software system that runs on the \dt and also lacks the ability of proper bidirectional synchronization between \dt and \dt.

\subsection{Conceptual Models to Define Digital Twins}\label{subsec:semiformal}
The presented research projects and papers leave plenty of space for interpretation of the digital twin concept. This is one reason, why there are so many definitions of digital twins.

\textcite{Yue2021} present a semi-formal approach using UML class diagrams to define the \pt, \dt and their relationships by the example of an automated warehouse system (AWS). \Cref{fig:awsuml} depicts the relationships. Physical twin and \dt exchange data via the \textit{PT-To-DT-Connection} and \textit{DT-To-PT-Connection}. A state change in one twin, triggers the change of the state of its counterpart.

Furthermore, they payed attention to two aspects, which are often not considered explicitly: fidelity and the twinning rate. Fidelity considers the accuracy and the level of abstraction of the \dt and the twinning rate is the interval \pt and \dt synchronize their states.

However, the semi-formal approach by \textcite{Yue2021} has its flaws. Although they considered the digital model as part of the \dt, it is not explicitly mentioned in the general overview in \Cref{fig:awsuml}. Moreover, the digital shadow was ignored completely.

\textcite{Becker2021} present in their conceptual model of digital shadows for CPS in a simlar approach using also UML class diagrams to show the relationships, but solely for the digital shadow. The focus of the digital shadow is on single assets and their information flow from the \pt to the digital shadow. They also emphasize that an asset's corresponding model is part of the digital shadow and models can be of different natures/types.

A formal mathematical approach, yet very abstract, of the relationships between physical twins, digital shadow, and digital twin was presented by \textcite{Lv2023}. A limitation in their approach is that it still offers a lot of space for interpretation and the mathematical notation is peculiar.

In this paper, we extend and merge the relationship diagrams of \textcite{Yue2021} and \textcite{Becker2021} by also including the digital model and digital shadow to give a full overview of the Digital Twin concept. In addition, we present the formalization of a \dt software architecture using the Object-Z notation.

\subsection{Continuous Twinning}\label{subsec:continuoustwinning}
In the development phase of CPS, \hil testing still is the common approach. The pressure to reduce costs \cite{SIsoftwarequality} led to many different approaches to switch from \hil to \sil. To date, for most industrial applications, sensors and actuators are connected via input/output ports to programmable logic controllers (PLCs). Although new wireless communication technologies and more powerful and efficient single-board computers open up the embedded community for cheaper and faster development processes, the predominance of PLCs will hold for years. It is quite common to use PLCs in a \hil setup, where the PLC is connected to a simulation \cite{Lyu2021}. Engineers can program the PLC and the simulation delivers the virtual context with simulated sensors/actuators to the PLC. As still only one engineer can work on a \hil system at the same time, \sil approaches become more and more popular to enable the collaboration between engineers. \textcite{Lyu2021} demonstrated that a software PLC in a \sil context can be realized with Docker and other tools.

Quality assurance of embedded systems is regulated with standards and norms to ensure robust testing and to prevent malfunctions that might pose a risk to the safety of individuals who work with or use these systems \cite{interviewpaper}. The aviation industry is renowned for its strict and stringent testing procedures, contributing to the fact that aircraft are the safest mode of transportation, statistically. This was not the case half a century ago, as standards and procedures have evolved through various experimentation with different testing strategies.

The \dtp approach presented in this paper, enables engineers to produce the first minimum viable product (MVP) with the first implemented device driver and emulator. Thanks to the publish-subscribe architecture, all additional nodes and emulators can be developed and added iteratively. Putting all modules in a source code management system allows all developers to use the \dtp and enhance the entire system incrementally, without the need to connect to the hardware of the \dt. As a bonus, this also enables automated SIL testing in continuous integration/continuous delivery (CI/CD) pipelines.

By following CI/CD workflows the development of embedded software systems becomes an agile and incremental process. Beginning with a prototype of a device driver for a single piece of hardware, to entire production plants, to smart factories, agile software development is enabled. This does not only improve the software quality and shorten release cycles, it also allows additional stakeholders to participate in a feedback loop in the development process from the first MVP. Adjusting software requirements or fixing design flaws can be done during development. With this method, digital twins evolve continuously in small incremental steps, rather than in major releases. \textcite{continuoustwinning} envision and call this approach \emph{Continuous Twinning}.

\section{The Digital Twin Concept - A Formalization}\label{sec:digitalconcept}
As \textcite{grieveskritzinger} elaborates, there is a flaw in the categorization of the \dt definition by \textcite{dtdef-kritzinger}. Stating that \dts have three subcategories, where a \dt is a subcategory of itself, leads to endless recursion. Furthermore, this increases the confusion around what a \dt is and what it is not. However, we do not share the recommendation to ignore the difference between a digital shadow and a digital twin with \textcite{grieveskritzinger}. To enhance clarity around the concepts and relationships between physical twins, digital models, digital shadows, digital threads, digital twin prototypes, digital templates, and digital twins, we formally specify the Digital Twin concept as follows. We propose, similar to \textcite{Hasselbring2015}, a three-level interleaving of formality in the specification: 

\begin{enumerate}
    \item informal prose explanation and illustrations with examples;
    \item semi-formal object-oriented modeling with the UML;
    \item rigorous formal specification with Object-Z.
\end{enumerate}

\begin{figure*}[b]
    \centering
    \includegraphics[width=0.8\textwidth]{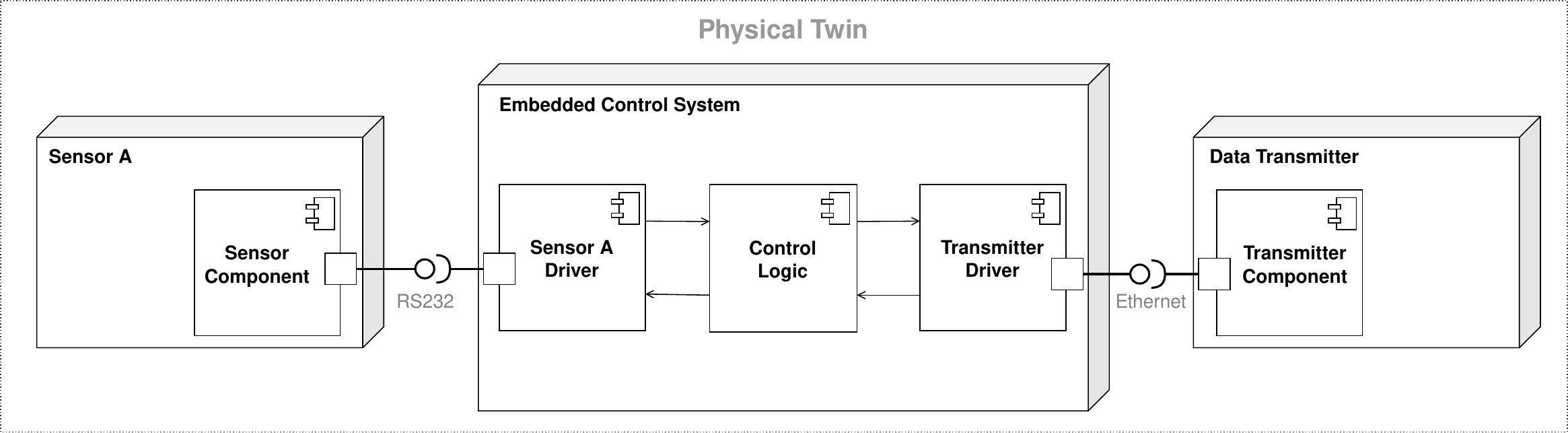}     
    \caption{The deployment diagram of an embedded system comprising a sensor, a data transmitter and the embedded control system 
    both are connected to. The sensor is connected via RS232 and the transmitter via transmitter via Ethernet.}
    \label{fig:wholesystem}
\end{figure*}
Object-Z \cite{smith2012object} is a formal specification notation used to describe the behavior of software systems. It extends the Z notation \cite{spivey1992z} and enables the incorporation of object-oriented concepts, such as classes, objects, inheritance, and polymorphism, into specifications. Additionally, Object-Z allows for the specification of operations that can be performed on objects, along with constraints on attribute values and relationships between objects, all expressed in a mathematical notation. The following specification has been checked using a type checker provided by the Community Z Tools Project \cite{czt}.

The formal specification is exemplified through an embedded software system comprising a sensor, an actuator which also serves as a data transmitter, and an embedded control system connected to both. This control system manages data and command exchange between these components. All example components are very basic and are only meant to demonstrate the core ideas. A real system would be more complex, including more third-party dependencies, tools, and frameworks.

\subsection{The Physical Twin}\label{sec:ptobjectz}
The digital twin concept starts with the physical twin.

\begin{tcolorbox}
\begin{definition}[Physical Twin]\label{def:pt}
A physical twin is a real-world physical System-of-Systems or product. It comprises sensing or actuation capabilities driven by embedded software.
\end{definition}
\end{tcolorbox}
\Cref{fig:wholesystem} illustrates the deployment diagram of our simple embedded system. In this example, the sensor is connected via an RS232 interface to the controller, and the transmitter is connected via Ethernet. All data collected from the sensor is processed by the controller logic and subsequently sent to an external source via the transmitter. Commands to modify the sensor's behavior are received by the transmitter and forwarded to the sensor through the control logic.

Consider both devices as black boxes that maintain a list of accepted commands, a method for executing tasks based on the commands and returning a result, and functions for sending and receiving data. Additionally, a device driver holds a corresponding list of commands that can be sent to the devices. The lists on the device and the device driver are identical, and the device driver handles command transmission and response reception.

The UML class diagram in \Cref{fig:umlall} depicts the various classes forming the embedded control system. To align with the clean code principles, abstract classes \textit{Device} and \textit{DeviceDriver} are introduced first. Sensors and actuators are considered as devices and thus inherit from \textit{Device}, as depicted on the left side of \Cref{fig:umlall}. All devices are connected to the embedded control system.

\begin{figure*}[ht]
    \centering
    \includegraphics[width=0.75\textwidth]{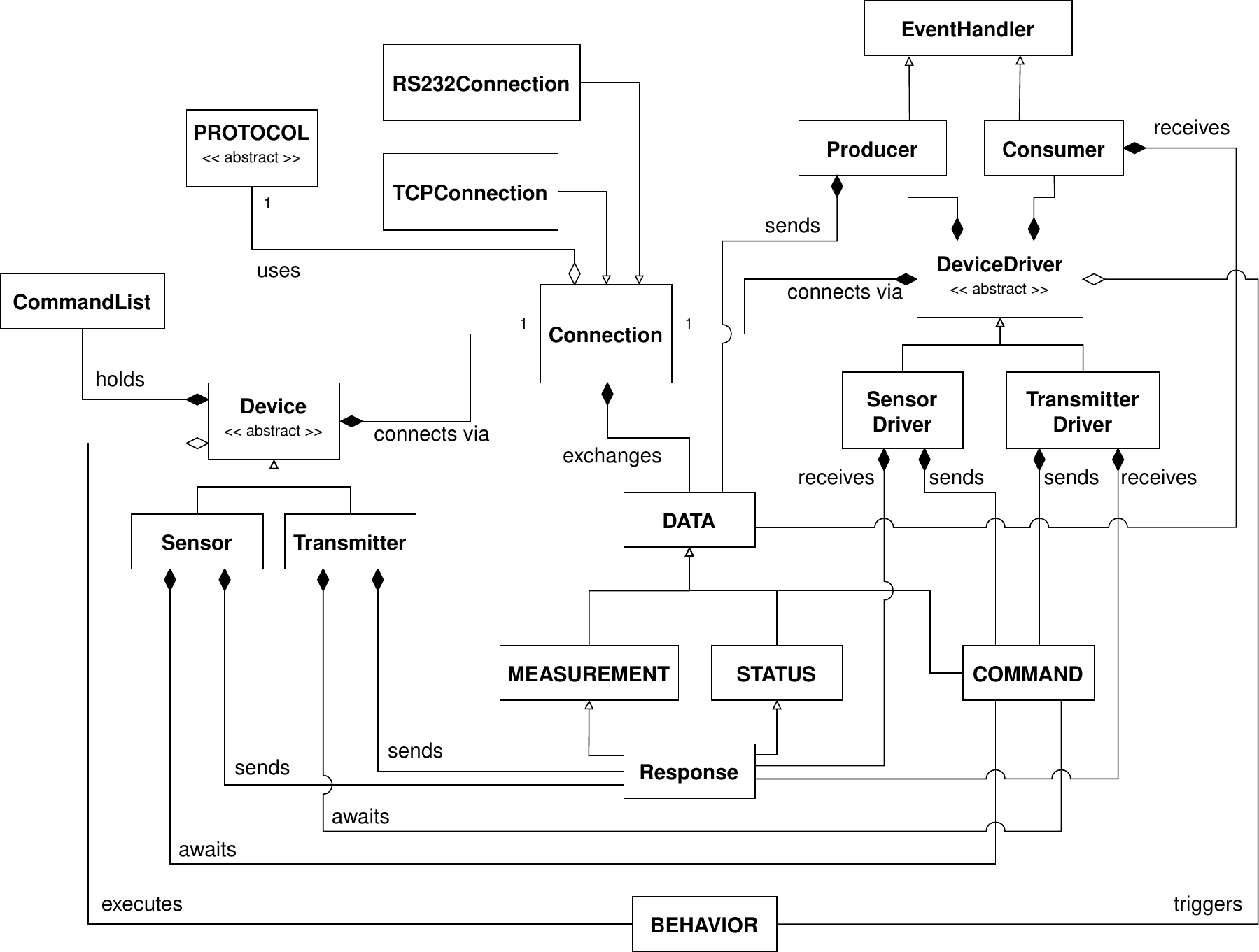} 
    \caption{UML class diagram of a \pt.}
    \label{fig:umlall}
\end{figure*}

The crucial elements of embedded software systems are the connections between the control systems and the sensors/actuators. In this example, the connections are established using different \textit{PROTOCOL} types (TCP or RS232) to facilitate communication between \textit{Device} and \textit{DeviceDriver}.

Specifically, \textit{SensorDriver} inherits from \textit{DeviceDriver} and employs an \textit{RS232Connection} to establish a connection with a \textit{Sensor}. Similarly, \textit{Transmitter} and \textit{TransmitterDriver} (which also inherits from \textit{DeviceDriver}) establish a connection using \textit{TCPConnection}. While a \textit{Device} is treated as an external component running on the device, a corresponding \textit{DeviceDriver} is an integral part of the embedded control system.

A \textit{Device} consists of two main components: a \textit{Connection} object and a set of accepted commands (\textit{commandList}). The \textit{Connection} object manages data exchange between a \textit{Device} and a \textit{DeviceDriver}. The \textit{ExecuteCommand} function represents the execution of a task after a command has been sent to the \textit{Device}. It expects a \textit{COMMAND} object sent by the \textit{DeviceDriver} and returns a \textit{RESPONSE} object. The \textit{Send} and \textit{Receive} functions utilize the corresponding functions provided by the contained \textit{Connection}.

To facilitate the exchange of data from a sensor to another process, such as the control logic, \textit{EventHandler} objects are introduced. It can be assumed that these \textit{EventHandler} objects are implemented in a manner similar to the Observer pattern, which also encompasses publish/subscribe architectures.

In this setup, all events received from the \textit{Sensor} are emitted to all listeners through a \textit{Producer}, and processes receive these events by including a \textit{Consumer}.

\subsubsection{Object-Z Formalization}
The specification of this simple embedded system follows a bottom-up approach. The deployment diagram, as depicted in \Cref{fig:wholesystem}, can be defined using the Object-Z notation. To achieve this, some basic type definitions are introduced:
\begin{zed}
[PROTOCOL, EVENT] \\
\end{zed}
\textit{PROTOCOL} represents the communication protocols utilized between the devices and the control system, while \textit{EVENT} is the type employed for data exchange between processes.

Basic type definitions introduce new types in Z and Object-Z. Such internal structure is considered irrelevant for the specification. In this particular specification, any details that are not architecturally relevant are abstracted this way.

The various \textit{PROTOCOL} types used in the schema architecture are subsequently defined through an axiomatic definition. In this context, \textit{TCP} and \textit{RS232} are established as values of type \textit{PROTOCOL}:

\begin{axdef}
    TCP, RS232: PROTOCOL \\
\end{axdef}

Up until this point, only basic types have been introduced. However, as Object-Z is object-oriented, objects are also created. In this context, the parent class is denoted as \textit{DATA}, and it will later be specialized through inheritance into classes specific to the various data types:

\begin{class}{DATA}
    \begin{state}
        data: \seq \{0,1\}
    \end{state}
\end{class}

Communication between devices is represented as a sequence of bits. Given that standard data types such as integers, floats, or strings are irrelevant for the specification, only a bit representation is utilized.

As both a device and its corresponding device driver exchange either \textit{RESPONSE} or \textit{COMMAND}, the corresponding schemas inherit from the \textit{DATA} class. In this context, \textit{RESPONSE} can represent either \textit{MEASUREMENT} or \textit{STATUS}:

\begin{sidebyside}
\begin{class}{COMMAND}
DATA \\
\end{class}
\nextside
\begin{class}{STATUS}
DATA \\
\end{class}
\end{sidebyside}

\begin{sidebyside}
\begin{class}{MEASUREMENT}
DATA \\
\end{class}

\nextside
\begin{class}{RESPONSE}
MEASUREMENT \\
STATUS
\end{class}
\end{sidebyside}

Once the data types have been formalized, the various components and their connections can be configured. Initially, the abstract \textit{Connection} class can be defined as follows: 
\begin{class}{Connection}
\project (\Init, Read, Write) \\
\begin{state}
    type: PROTOCOL  \\
    dataStream: \seq DATA \\
\end{state}\\  \zbreak
\begin{init}
    dataStream = \langle \rangle \\
\end{init} \\  \zbreak
\begin{op}{Write}
    \Delta (dataStream) \\
    value?: DATA
    \where
    dataStream' = dataStream \cat \langle value? \rangle
\end{op} \\   \zbreak
\begin{op}{Read}
    \Delta (dataStream) \\
    value!: DATA
    \where
    dataStream = \langle value! \rangle \cat dataStream'
\end{op} \\   \zbreak
\begin{classcom}
    \textnormal{The symbol ? denotes input parameters and ! denotes outputs \cite{smith2012object}.}
\end{classcom} \\  
\end{class}
A \textit{Connection} possesses a type and manages bit sequences, represented as a stream (\textit{dataStream}). The \textit{Write} function appends bit sequences to the stream, while the \textit{Read} function extracts them by reading bits from it.

The specific implementations, \textit{RS232Connection} and \textit{TCPConnection}, are named after 
the types they set for the \textit{Connection} object from which they inherit:

\begin{sidebyside}
\begin{class}{RS232Connection}
Connection\\
\begin{state}
    type = RS232  \\
\end{state}\\
\end{class}
\nextside
\begin{class}{TCPConnection}
Connection\\
\begin{state}
    type = TCP  \\
\end{state}\\
\end{class}
\end{sidebyside}

A \textit{Device} comprises a \textit{Connection} object and a set of accepted commands (\textit{commandList}). The \textit{Connection} object is responsible for managing data exchange between a \textit{Device} and a \textit{DeviceDriver}. The \textit{ExecuteCommand} function represents the execution of a task following the transmission of a command to the \textit{Device}. It expects a \textit{COMMAND} object sent by the \textit{DeviceDriver} and returns a \textit{RESPONSE} object. The \textit{Read} and \textit{Write} functions make use of the corresponding functions provided by the contained \textit{Connection}:

\begin{class}{Device}
\project (\Init, Send, Receive, commandList) \\
\begin{state}
    connection: \poly Connection \copyright\\
    commandList: \power COMMAND \\
    \where
    connection \notin Connection \\
    \# commandList > 0
\end{state}\\  \zbreak
\begin{classcom}
    \textnormal{The symbol $\poly$ denotes the union of Connection with all sub-types. Connection is abstract, thus the Connection has to be sub-type 
    that implements it. \\
    The symbol $\copyright$ denotes object containment \cite{smith2012object}.}
\end{classcom} \\  \zbreak
\begin{init}
    connection.\Init \\
\end{init} \\  \zbreak
\begin{op}{ExecuteCommand}
    command?: COMMAND \\
    result!: \poly DATA
    \where
    command? \in commandList
\end{op}\\  \zbreak
Send \sdef connection.Write \\
Receive \sdef connection.Read \semi ExecuteCommand \\
 \qquad \quad \semi Send \\ \zbreak
 \begin{classcom}
     \textnormal{The symbol $\semi$ denotes a sequential composition.}
 \end{classcom}
\end{class}

Similar to the \textit{Device} class, the \textit{DeviceDriver} class also contains a \textit{Connection} object, a set of commands, a set of known behaviors, and a function that maps a behavior to the corresponding command that can be sent to the \textit{Device}:
\begin{class}{DeviceDriver}
\project (\Init, Send, Receive, commandList, emitter, \\ consumer) \\
\begin{state}
    connection: \poly Connection \copyright\\
    commandList: \power COMMAND \\
    emitter: Producer \copyright\\
    consumer: Consumer \copyright\\
        \where
    connection \notin Connection \\
\end{state} \\  \zbreak
\begin{init}
    connection.\Init \land consumer.\Init \\
\end{init} \\  \zbreak
Send \sdef consumer.Consume \semi connection.Write \\
Receive \sdef connection.Read \semi emitter.Emit
\end{class}

Assume for this example that the \textit{DeviceDriver} fully implements all interactions with the \textit{Device} and hence, the \textit{commandList} for both instances is equal.
The \textit{Receive} and \textit{Send} functions in this class also utilize the \textit{Connection}'s \textit{Read} and \textit{Write} functions. Any further implementations beyond this scope are not relevant to our specification.

Data exchange between different processes, such as the \textit{DeviceDriver} and the \textit{ControlLogic}, occurs through \textit{EventHandler}s: 

\begin{class}{EventHandler}
\project (event) \\
    \begin{state}
        event: EVENT
    \end{state} \zbreak
\end{class}

Each \textit{EventHandler} registers for a specific \textit{EVENT}, which can represent, for example, a simple response from the \textit{Device}. In this example, the \textit{EventHandler} is an abstract class, and \textit{Producer} and \textit{Consumer} are the specific implementations. Assuming both register for the same \textit{EVENT}, like ``NEWDATA,'' a \textit{Producer} can emit new events, and the \textit{Consumer} receives and handles all incoming events. It is important to note that this relationship is not one-to-one but rather one-to-many, allowing for an indefinite number of \textit{Consumer}s to listen to the same \textit{Producer}.

The main function of a \textit{Producer} is the \textit{Emit} function that is called with a passed \textit{DATA} object
and then all \textit{Consumer}s are notified:

\begin{class}{Producer}
\project (\Init, event,  Emit) \\
EventHandler\\
    \begin{op}{Emit}
        occuredEvent?: \poly DATA \\
        eventToEmit!: \poly DATA 
        \where
        eventToEmit! = occuredEvent?
    \end{op}\\  \zbreak
\end{class}

A \textit{Consumer} registers via the \textit{Observe} to an \textit{EVENT} and only listens to the emitted events and handles them in a queue. The \textit{Consume} function returns
always the first element in the queue:

\begin{class}{Consumer}
\project (\Init, event, queue, Observe, Consume) \\
EventHandler\\
    \begin{state}
        queue: \power \poly DATA
    \end{state} \\ \zbreak
    \begin{init}
        queue = \emptyset
    \end{init} \\ \zbreak
    \begin{op}{Observe}
        \Delta (queue) \\
        item?: \poly DATA
        \where
        queue' = queue \cup \{ item? \}
    \end{op}\\  \zbreak
    \begin{op}{Consume}
        \Delta (queue) \\
         item!: \poly DATA
        \where
        \# queue > 0 \\
        item! \in queue \\
        queue' = queue \setminus \{ item! \} 
    \end{op} \\  \zbreak
\end{class}

After introducing the basic classes, the logic of the embedded control system can be defined. The \textit{DeviceDriver} manages all communication between the control system and the \textit{Device}, with communication being established through the \textit{Connection} class. In this scenario, assume this \textit{DeviceDriver} is straightforward and serves as a relay between the control logic and the device.

The \textit{Consumer} handles all incoming \textit{DATA} from the control logic and forwards them to the device. When responses are received from the device, the \textit{emitter} forwards these responses to all listeners.

In Object-Z, the symbol $\parallel$ represents a sequential execution. Therefore, the \textit{Send} function first receives an incoming event by invoking \textit{consumer.Consume}, and only afterwards, that call's result is received, it is passed to the \textit{Connection}, which then sends the command to the device. Conversely, incoming responses from the device are received from the connection using \textit{connection.Read} and subsequently emitted to all listeners through \textit{emitter.Emit}.

Now that the abstract classes for \textit{Device}, \textit{Connection}, and \textit{DeviceDriver} have been established, we can proceed to define the concrete classes for the sensor, named \textit{Sensor}, and its corresponding device driver, \textit{SensorDriver}, as depicted in Figure 3a. In this particular example, \textit{Sensor} and \textit{SensorDriver} are interconnected using an \textit{RS232Connection}.

The outcome of an executed command is categorized as a \textit{RESPONSE}, which can represent either a \textit{MEASUREMENT} or a \textit{STATUS} object. The remaining functions within these specific classes remain consistent with those in the abstract parent classes \textit{Device} and \textit{DeviceDriver}:

\begin{class}{Sensor}
\project (\Init, Send, Receive, commandList) \\
Device\\
\begin{state}
    connection: RS232Connection \copyright
\end{state} \\  \zbreak
\begin{op}{ExecuteCommand}
    command?: COMMAND \\
    result!: RESPONSE
    \where
    command? \in commandList
\end{op}\\  \zbreak
\end{class}
A \textit{SensorDriver} inherits the \textit{EventHandler}s from its parent class: 

\begin{class}{SensorDriver}
\project (\Init, Send, Receive, commandList) \\
DeviceDriver\\
\begin{state}
    connection: RS232Connection \copyright
\end{state} \\  \zbreak
\end{class}

In this example, all incoming commands are dispatched by the control logic, consumed by the driver, and subsequently forwarded to the sensor via the connection. Vice versa, all responses from the sensor are emitted as events by the corresponding producer and can be listened to by all consumers.

The essence of this specification lies in the communication between a device and its device driver, which is captured by the \textit{Communication} schema. In this instance, the device is a \textit{Sensor}, and the driver is a \textit{SensorDriver}. Both the device and the driver share the same \textit{commandsList} and are connected through an \textit{RS232Connection}.

In Object-Z, the symbol ``$\parallel$'' signifies the execution of functions in parallel \cite{smith2012object}. Therefore, \textit{ReadFromDevice} illustrates the \textit{Sensor} sending data while the corresponding \textit{SensorDriver} reads it. Conversely, \textit{ReadFromDriver} represents the reverse scenario, with communication from the \textit{SensorDriver} to the \textit{Sensor}:
\begin{class}{Communication}
\begin{axdef}
    device: Sensor \\
    driver: SensorDriver 
    \where
    \forall x: device.commandList \\ \qquad \spot x \in driver.commandList \\
    \forall x: driver.commandList \\ \qquad \spot x \in device.commandList
\end{axdef}\\  \zbreak
ReadFromDevice \sdef device.Send \parallel driver.Receive \\
ReadFromDriver \sdef driver.Send \parallel device.Receive \\ \zbreak
\begin{classcom}
\end{classcom}
\end{class}
The \textit{Transmitter} class is akin to the \textit{Sensor} class in many ways. It handles incoming commands and provides responses in return. However, since the \textit{Transmitter} is an actuator, it does not return measurements but instead sends data using another communication protocol, such as LoRaWAN. It is important to note that this communication differs from the \textit{Communication} schema described earlier. Additionally, the \textit{Connection} object solely represents the connection between the \textit{Device} and \textit{DeviceDriver} and does not pertain to the communication between two transmitters:

\begin{class}{Transmitter}
\project (\Init, Send, Receive, commandList) \\
Device\\
\begin{state}
    connection: TCPConnection \copyright
\end{state} \\  \zbreak
\begin{op}{ExecuteCommand}
    command?: COMMAND \\
    result!: RESPONSE
    \where
    command? \in commandList
\end{op}\\  \zbreak
\end{class}

Similar to the \textit{SensorDriver}, the \textit{TransmitterDriver} represents only a data relay between device and control logic:

\begin{class}{TransmitterDriver}
\project (\Init, Send, Receive, commandList) \\
DeviceDriver\\
\begin{state}
    connection: TCPConnection \copyright
\end{state} \\  \zbreak
\end{class}

The details of the control system are not within the scope of this specification. The control logic for an embedded system is often some form of a state machine. State machines fully automate a system, but do not adapt to new or changed processes on the fly. Modern Industry 4.0 application incorporate autonomous behavior, extracted or learned from gathered data and thus, include architectures different from state machines. Furthermore, the orchestration of processes, including different commands to different sensor and actuators, can be quite complex. However, for this example, the only function of the \textit{ControlLogic} class is to execute the commands received from the transmitter and return the responses from the sensor:

\begin{class}{ControlLogic}
\begin{state}
    sensor: Consumer \copyright \\
    transmitter: Consumer \copyright \\
    response: Producer \copyright \\
    command: Producer \copyright \\
    period: \num \\
    dataLog: \power\poly DATA \\
\end{state} \\\zbreak
\begin{init}
                period = 0 \\
                dataLog = \emptyset\\
                sensor.\Init\land transmitter.\Init
\end{init} \\ \zbreak
\begin{op}{SetPeriod}
    \Delta(period) \\
    newPeriod?: \num
    \where
    period' = newPeriod?
\end{op} \\ \zbreak
\begin{op}{LogData}
    \Delta(dataLog) \\
    newSensorData?: \poly DATA
    \where
    period > 0 \\
    dataLog' = dataLog \cup \{ newSensorData? \}
\end{op} \\ \zbreak
    sendCmd \sdef \semi data : transmitter.queue \\ \qquad \quad \spot transmitter.Consume \semi SetPeriod \\ \qquad \quad \semi command.Emit \parallel LogData \\
    sendRsp \sdef \semi data : sensor.queue \\ \qquad \quad \spot sensor.Consume \semi response.Emit \\ \qquad \quad \parallel LogData
\end{class}

The incoming commands contain the value that sets the sample rate of the sensor. To configure the period, the function \textit{sendCmd} processes events sequentially from the transmitter queue. For each event, the \textit{SetPeriod} function is called to set the sample rate. The newly configured period is then sent as a command to the sensor, which adjusts its sample rate accordingly. This message exchange is logged in a list called \textit{dataLog}.

Assume the commands from the transmitter only include a period for the sensor's sample rate. To configure the period, the function \textit{sendCmd} processes events sequentially from the transmitter queue. For each event, the \textit{ChangeBehavior} executes \textit{SetPeriod} to internally set the sample rate and newly configured period is then sent as a command to the sensor, which adjusts its sample rate accordingly. This message exchange is logged in a list called \textit{dataLog}.

All events originating from the sensor are handled by \textit{sendRsp} and are sent to the transmitter without any alterations. Once again, the message exchange is recorded in the data list through the \textit{LogData} command.

With all required classes defined, the schema of the \textit{EmbeddedControlSystem} from \Cref{fig:wholesystem} can be defined: 
\begin{class}{EmbeddedControlSystem}
\begin{state}
    sensorDriver: DeviceDriver \copyright \\
    transmitterDriver: DeviceDriver \copyright \\
    controlLogic: ControlLogic \copyright \\
\end{state} \\\zbreak
\end{class}
And finally, the union of the devices and the schema \textit{EmbeddedControlSystem} forms the \textit{Physical Twin}:
\begin{class}{Physical Twin}
\begin{state}
    sensor: Device \copyright \\
    transmitter: Device \copyright \\
    ecs: EmbeddedControlSystem \copyright \\
\end{state} \\\zbreak
\end{class}

\subsection{The Digital Model}
Modeling and simulation are powerful methods utilized in various fields to evaluate complex systems, processes, and knowledge. They empower researchers, engineers, and decision-makers to examine real-world phenomena within controlled and virtual environments. This, in turn, enables them to make informed decisions and gain insights into the system under investigation. At the core of modeling lies the concept of mathematical modeling, which plays a pivotal role in formally capturing the essence of the system.

Mathematical models are representations of real-world systems employing mathematical equations, relationships, and logical structures. They provide a means to describe and quantify the behavior of a system. While mathematical models are not confined to any specific domain, in this work, we concentrate on their application in the engineering domain.

Before the advent of computers, the construction of machines was primarily carried out on drawing boards. This paradigm shifted with the introduction of computer-aided designs (CAD), enabling the creation of 2D and 3D models that could be easily shared and replicated with others. Over the past decades, advancements in tooling and computational power have facilitated the substitution of real prototypes with virtual prototypes. This transition has significantly reduced design cycles and lowered design costs. When components of a system are governed by mathematical relationships, virtual prototypes can be rigorously tested in simulations across a wide range of conditions. This allows for the evaluation of potential design weaknesses, providing immediate feedback on design decisions.

The \textit{Digital Model} serves as a central component of a \dt. However, most definitions merely mention digital models, assuming that researchers share a common understanding of what a model entails. This often leads to the assumption that a CAD model constitutes the entirety of a digital model, while a simulation is considered something more than a digital model, despite both being forms of mathematical models. Hence, we define a digital model as follows:

\begin{tcolorbox}
\begin{definition}[Digital Model]
A digital model describes an object, a process, or a complex aggregation. The description is either a mathematical or a computer-aided design (CAD).
\end{definition}
\end{tcolorbox}
This definition encompasses various aspects of digital modeling, including the use of CAD as the foundational model for system design, its utilization within simulation tools involving complex processes, and even purely mathematical models.

\subsubsection{Introducing the State Machine Example}
Although the \pt is defined as including (autonomous) behaviors instead of a state machine, this example could also be implemented as a state machine, where one can model its different states as follows:
\begin{equation}\label{eq:statemachine}
\begin{aligned}
\mathcal{M} = (Q, \textstyle\sum, \delta, q_0, F)
\end{aligned}
\end{equation}
A state machine $\mathcal{M}$ can be represented by a 5-tuple $M$, which consists of a finite set of states $Q$, a finite set of input symbols known as the alphabet $\textstyle \sum$, a transition function $delta$ defined as  $\delta : Q \times \sum \to Q$, an initial or starting state $q_0 \in Q$, and a set of accept 
states $F\subseteq Q$.
The creation of state machines, often done using tools like LabView, remains a common approach employed by engineers for programming machines. This practice falls within the scope of the provided definition of a digital model.

The state machine of the embedded control system can be defined as follows:
\begin{itemize}
    \item $Q= \{ STANDBY, ACTIVE, OFF\}$
    \item $q_0 = STANDBY$
    \item $\sum = \mathbb{Z}$
    \item $\delta: Q \times \sum \to Q$ 
\end{itemize}
The corresponding UML state diagram is presented in \Cref{fig:statemachine}. Upon initiation, the initial state is \textit{STANDBY}, with the corresponding period value for the sensor's sampler rate set to 0, indicating that no samples are taken at this point. If a command with a value $x \in \sum$, where $x > 0$, is issued, the state machine transitions to the \textit{ACTIVE} state. Conversely, if a command with a value $x = 0$ is received, the state reverts to \textit{STANDBY}. For values of $x < 0$, the state of the system changes to \textit{OFF}.

\begin{figure}[ht]
    \centering
    \includegraphics[width=0.45\textwidth]{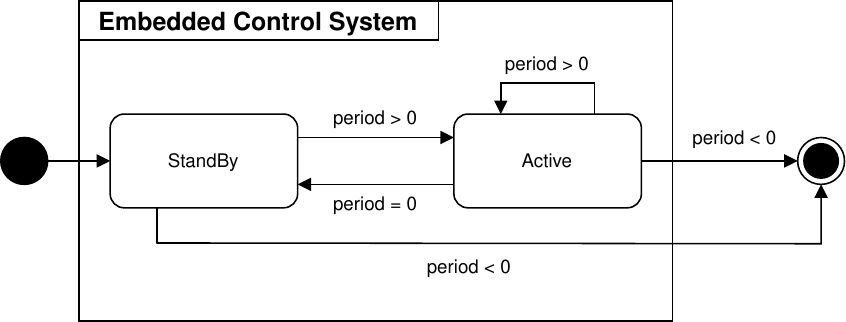}   
    \caption{A state machine of the embedded control system formalized for the \pt.}
    \label{fig:statemachine}
\end{figure}

\begin{figure}[ht]
    \centering
    \includegraphics[width=0.45\textwidth]{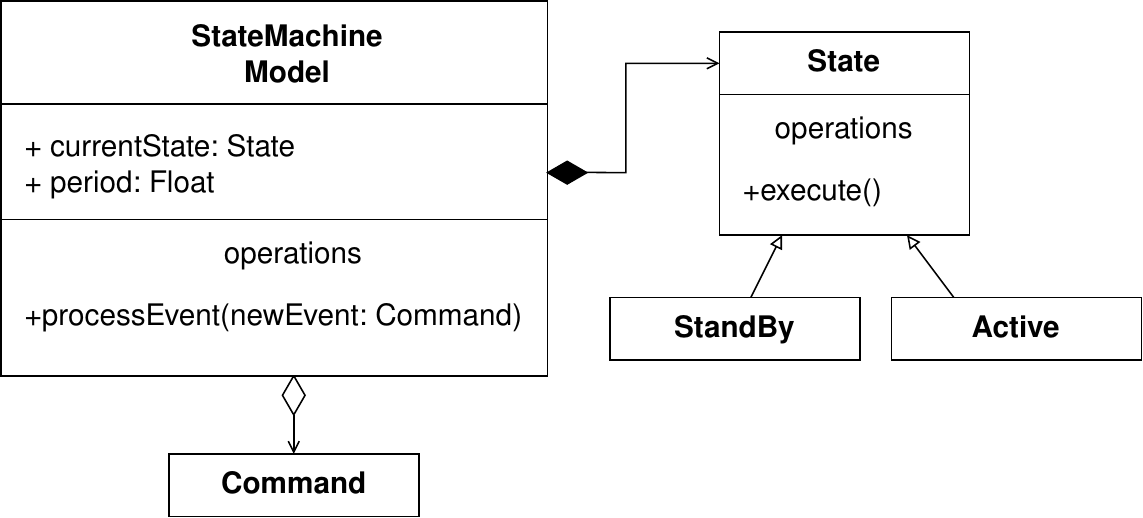}  
    \caption{UML class diagram for the state machine.}
    \label{fig:statemachinecd}
\end{figure}

\subsubsection{Object-Z Formalization}
This state machine can also be specified in Object-Z. First, the class diagram is displayed in \Cref{fig:statemachinecd}. \textit{STATE} is the parent class:
\begin{class}{STATE}
\project (execute) \\
    \begin{op}{execute}
    \end{op}
\end{class}

The \textit{execute} method will be internally overwritten by the child states. For this example, the specific code that is executed is irrelevant. The states the state machine can be in are defined as subclasses:
\begin{sidebyside}
    \begin{class}{ACTIVE}
        STATE
    \end{class}
    \\
    \begin{class}{OFF}
        STATE
    \end{class}
    \nextside
    \begin{class}{STANDBY}
        STATE
    \end{class}
\end{sidebyside}

The \textit{EventStateMachine} encapsulates the logic responsible for state changes upon receiving \textit{COMMAND} events and maintains both a \textit{STATE} (\textit{state} is also the variable) and a \textit{period}, which is a number. Initially, the \textit{period} is set to 0, corresponding to the initial \textit{state} set as \textit{STANDBY}. The \textit{ProcessEvent} function is responsible for modifying the state of the state machine in response to incoming events.
\begin{class}{EventStateMachine}
\project (\Init, ProcessEvent, state) \\
\begin{state}
    state:  \poly STATE \\
    period: \num
\end{state}\\  \zbreak
\begin{init}
    period = 0
\end{init} \\  \zbreak
\begin{op}{ProcessEvent}
\Delta(state) \\
    newEvent?: COMMAND \\
    newState!: \poly STATE
    \where
    state' = newState!
\end{op}
\end{class}

It is important to note that, at this stage, the \textit{EventStateMachine} has no connection to the \pt. All modifications and updates are made manually, and there is no automatic synchronization between the digital model and \pt. The schema for the digital model than includes the state machine:
\begin{class}{DigitalModel}
\project (\Init, ProcessEvent) \\
\begin{state}
    stateMachine: EventStateMachine \copyright    
\end{state} \\ \zbreak
\begin{init}
    stateMachine.\Init \\
\end{init} \\  \zbreak
ProcessEvent \sdef stateMachine.ProcessEvent
\end{class}

\subsection{The Digital Template}
In their initial definition of \dts, \textcite{dtdef-grieves} view the \dt as a collection of information necessary for constructing and monitoring the physical object. Specifically, the \dtp can be regarded as a virtualized set of blueprints, bills of materials, technical manuals, and similar documentation. When combined with the digital model, which can be used to extract all the information needed for creating blueprints and bills of materials, it can indeed be employed to construct and maintain the \pt

\begin{figure*}[ht]
    \centering
    \includegraphics[width=.8\textwidth]{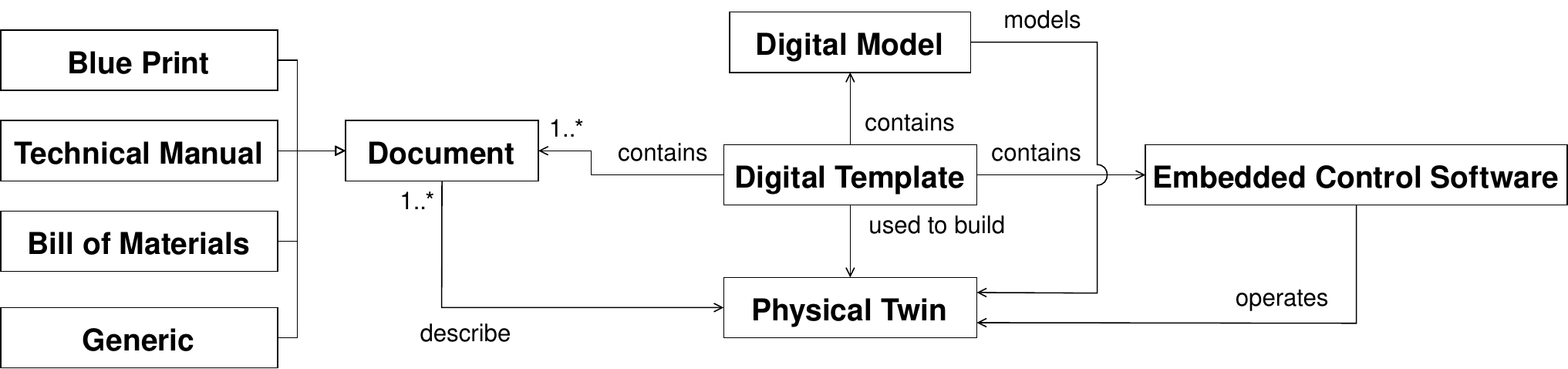}
    \caption{Digital Template}
    \label{fig:digitaltemplate}
\end{figure*}

However, this approach does not completely virtualize the \pt, as later demonstrated by the example of the OSI Model in \Cref{fig:isoosi} on Page~\pageref{fig:isoosi}. Thus, the early interpretation of this definition does not fully realize a \dt of a \pt.

To encompass all available materials for constructing and maintaining the \pt, including the software running the \pt and the digital model, these components can be bundled together into a comprehensive package. We refer to this bundle as the Digital Template.

\begin{tcolorbox}
\begin{definition}[Digital Template]
A digital template serves as a framework that can be tailored or populated with specific information to generate the \pt. It encompasses the software operating the \pt, its digital model, and all the essential information needed for constructing and sustaining the \pt, such as blueprints, bills of materials, technical manuals, and similar documentation.
\end{definition}
\end{tcolorbox}

\textcite{dtdef-grieves} initially defined digital template as a \dtp. However, in \textcite{grieves2023}, they expanded upon their definition of a \dtp. Their \dtp is all the products that can be made, including all their variants. They take shape over time, from an idea to a first manufactured article \cite{grieves2023}. We still consider that early versions of their \dtp are only a digital template. However, fully developed, they could also include the \dtp definition presented later in this work.

\subsubsection{Object-Z Formalization}
The UML class diagram of a digital template is depicted in \Cref{fig:digitaltemplate}. The digital template includes all documents that either describe the \pt or are required to build it. Furthermore, it includes the digital model the real system is derived from and the software that operates the \pt later.
For an Object-Z formalization, the general class \textit{Document} is defined:
\begin{class}{Document}
\end{class}
Specific types inherit from the \textit{Document} class:
\begin{sidebyside}
\begin{class}{BluePrint}
Document \\
\end{class}
\nextside
\begin{class}{TechnicalManual}
Document \\
\end{class}
\end{sidebyside}

\begin{sidebyside}
\begin{class}{BillOfMaterials}
Document \\
\end{class}
\nextside
\begin{class}{Generic}
Document \\
\end{class}
\end{sidebyside}
The schema for the digital template includes all the documents, the embedded control software and the digital model:
\begin{class}{DigitalTemplate}
    \begin{state}
        documents: \power Document \\
        ecs: EmbeddedControlSystem \\
        digitalModel: DigitalModel
    \end{state}
\end{class}

\subsection{The Digital Thread}
With the development of CPS, machines began interacting with servers tasked with monitoring and controlling them. This paradigm also applies to \dts. In this context, the communication channel facilitating such interaction is referred to as a digital thread. Taking inspiration from \textcite{leiva2016digitalthread}, we define the digital thread as follows:

\begin{tcolorbox}
\begin{definition}[Digital Thread]\label{def:digitalthread}
The digital thread refers to the communication framework that allows a connected data flow and integrated view of the \pt's data 
and operations throughout its life-cycle.
\end{definition}
\end{tcolorbox}
Data accumulated from physical objects can only be preserved if these objects possess an interface for storing the generated data. Similar to the general \dt definitions, there is, currently, no universally accepted and standardized solution for digital threads, given their diverse applications across various domains.

\begin{figure*}[bh]
    \centering
    \includegraphics[width=.7\textwidth]{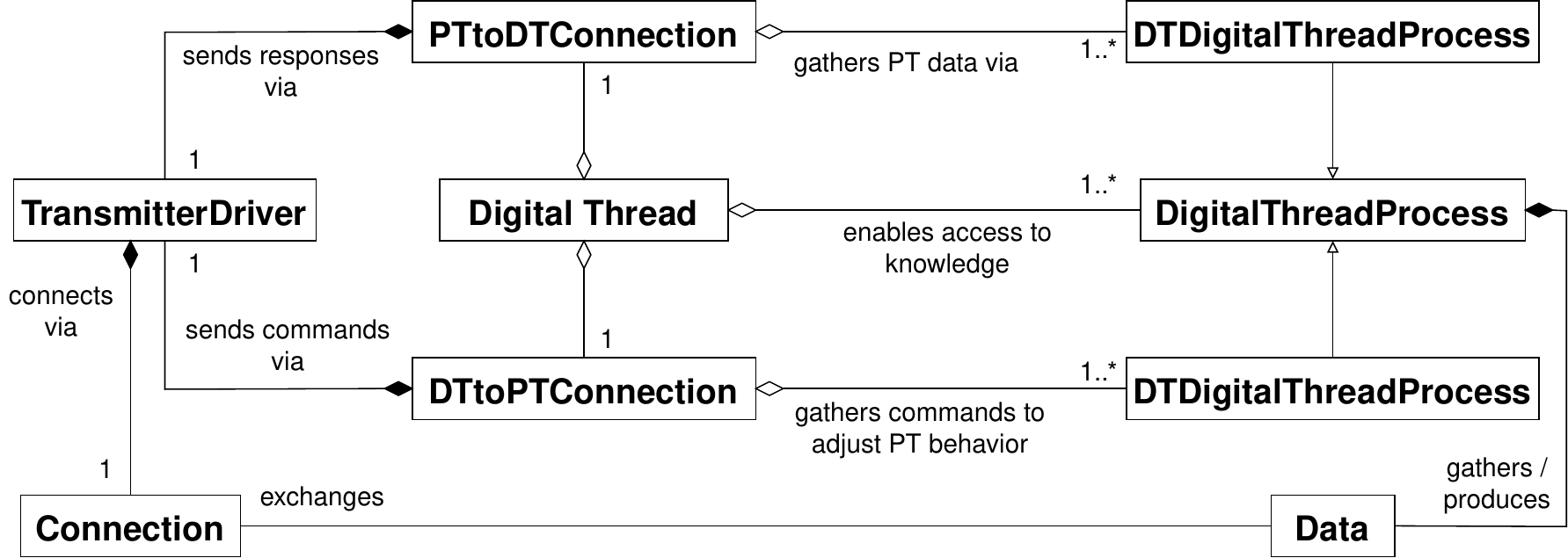}
    \caption{Class diagram of a digital thread.}
    \label{fig:dthreaduml}
\end{figure*}

Furthermore, it is crucial to understand that the digital thread encompasses more than just the communication protocol. It also involves applications and functionalities that assist in tasks such as monitoring, analysis, planning, and execution. These applications have the capacity to incorporate and share knowledge derived from the digital template and the gathered data preserving the \pt's evolution through time \cite{dtdef-saracco}.

\subsubsection{Object-Z Formalization}
The UML class diagram for a digital thread between the previously formalized \pt and a \dt, which will be defined later in this paper, is illustrated in \Cref{fig:dthreaduml}. The \textit{DigitalThread} exists of a \textit{PTtoDTConnection} that sends measurement and status messages (see the \textit{RESPONSES} Object-Z class) and the \textit{DTtoPTConnection}, which sends commands to the \pt. To send data, a \textit{TransmitterDriver} is used to to establish a \textit{Connection}. Notice that this connection is not between a \textit{DeviceDriver} and a \textit{Device}, but between two transmitters, e.g. using the LoRaWAN protocol. Both connection types gather data from processes (\textit{DigitalThreadProcess}). In general, these processes can be different in each digital thread. Referencing our example again, the \textit{ControlLogic} represents a \textit{PTDigitalThreadProcess}, since it forwards all sensor message to the transmitter, which then can transmit the data to the \dt. On the \dt's side, the \textit{DTDigitalThreadProcesses} can include many different kinds of processes. However, there are is at least one process that is included: the process that decides which command is sent to the \pt to adjust its sample rate. Since the digital thread is meant to show the evolution of the \pt over its life-cycle, all the gathered data has to be stored in some form of a database. Hence the database is a \textit{DigitalThreadProcess} that is part of the digital thread.

Formalizing this with Object-Z, we first define the \textit{DigitalThreadProcess}:
\begin{class}{DigitalThreadProcess}
\begin{state}
    knowledge: \power DATA
\end{state} \zbreak
\end{class}
A \textit{DigitalThreadProcess} has a set of \textit{DATA} that can be shared with the corresponding twin counterpart. \textit{PTDigitalThreadProcess} and \textit{DTDigitalThreadProcess} are derived classes:
\begin{sidebyside}
\begin{class}{PTDigitalThreadProcess}
DigitalThreadProcess
\end{class}
\nextside
\begin{class}{DTDigitalThreadProcess}
DigitalThreadProcess
\end{class}
\end{sidebyside}

This data is sent via the \textit{Connection} in the \textit{PTtoDTConnection}, which is defined as follows:
\begin{class}{PTtoDTConnection}
\begin{state}
    connection: TransmitterDriver \copyright \\
    collectFrom: \power PTDigitalThreadProcess
\end{state}
\end{class}
The counterpart is the \textit{DTtoPTConnection}:
\begin{class}{DTtoPTConnection}
\begin{state}
    connection: TransmitterDriver \copyright \\
    collectFrom: \power DTDigitalThreadProcess
\end{state}
\end{class}
Both classes form the \textit{DigitalThread}:
\begin{class}{DigitalThread}
\begin{state}
    ptData: PTtoDTConnection \\
    dtData: DTtoPTConnection \\
\end{state}
\end{class}

\subsubsection{The MAPE-K Reference Model}
To illustrate this concept more concretely, we demonstrate in the following section how the digital thread can serve as the connection between different phases within the MAPE-K (Monitor-Analyze-Plan-Execute over a shared Knowledge) reference model, as depicted in \Cref{fig:mapek}. MAPE-K extends IBM's MAPE framework, with ``K'' signifying ``Knowledge.'' The MAPE-K reference model provides a framework for automation of processes and the control loop for managing and optimizing computer systems.
The different stages, Monitor, Analyze, Plan, and Execute represent a specific stage or function within an autonomic computing system: 

\begin{figure}[ht]
    \centering
    \includegraphics[width=0.45\textwidth]{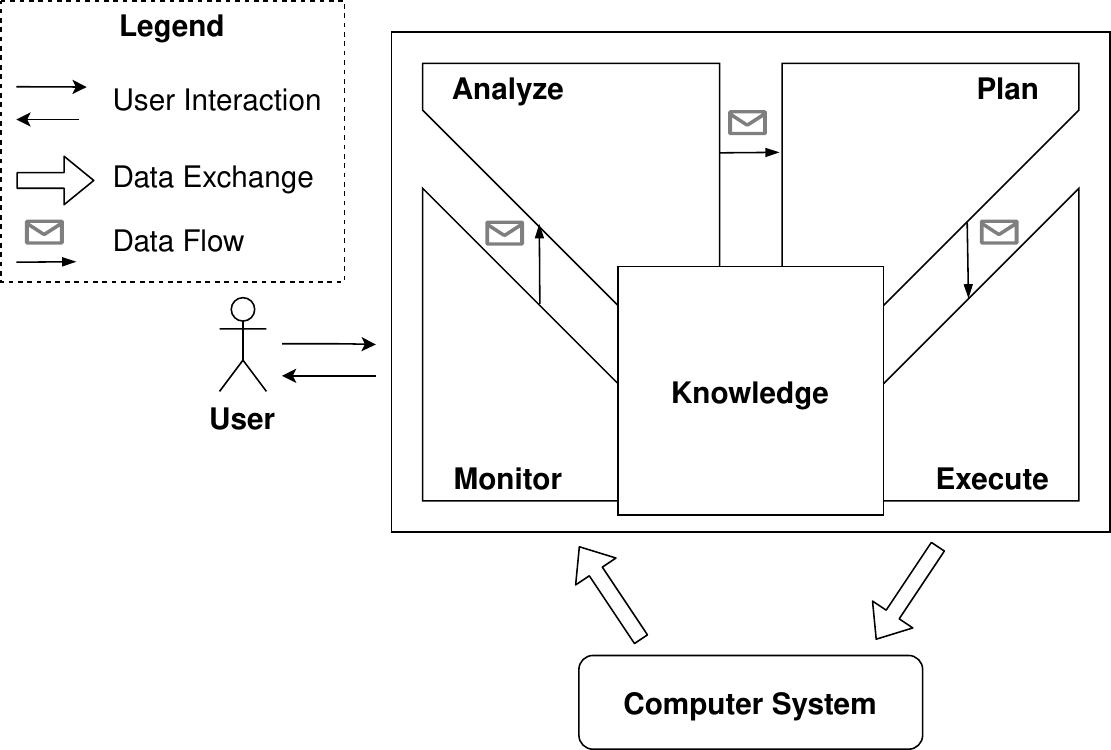}
    \caption{MAPE-K reference model for cyber-physical system.}
    \label{fig:mapek}
\end{figure}

\begin{itemize}
\item \textbf{Monitor}: This is the first stage of the framework. In this phase, the system continuously collects data and monitors its own performance and the surrounding environment. This can involve data from various sensors, actuators, or monitoring tools that gather information about the system's behavior, resource utilization, and external conditions.
\item \textbf{Analyze}: To gain insights into the system's behavior and performance, the data collected through monitoring, gets analyzed. The goal is to identify patterns, anomalies, and potential issues and hence, to understand the current state of the system.
\item \textbf{Plan}: Based on the analysis of the system's current state, the system formulates a plan for actions to be taken. This plan may involve adjustments, optimizations, or corrective measures aimed at improving system performance, resource allocation, or other relevant parameters. 
\item \textbf{Execute}: In the last phase, the system carries out the actions defined in the planning stage. These actions can be automatic or semi-automatic, depending on the level of autonomy and control designed into the system. The system implements the planned changes to achieve the desired state.
\item \textbf{Knowledge:} This component is critical for learning and adaptation. It involves maintaining a repository of historical data, models, policies, and best practices. The system uses this knowledge to make more informed decisions in subsequent iterations of the MAPE-K loop. Over time, the system becomes better at self-optimization and self-management by learning from its past experiences.
\end{itemize}

These stages are executed sequentially one after another and all have permanent 
access to the Knowledge about the system. The realization of the data flow between the different stages is part of the Digital Thread.
Also, applications around the different stages, which are, for instance, connected via APIs, are also part of the Digital Thread,
if they provide better insight for the corresponding \pt to the user.

\subsection{The Digital Shadow}
To fully harness the potential of the digital thread, a process situated at either end of the digital thread must consolidate all the disparate elements into a platform that users can utilize to gain insights into the current state of the \pt. In the context of the Digital twin concept, this role is fulfilled by the digital shadow. The digital shadow is defined as follows:

\begin{tcolorbox}
\begin{definition}[Digital Shadow]
A digital shadow is the sum of all the data that are gathered by an embedded system from sensing, processing, or actuating. The connection from a \pt to its digital shadow is automated. Changes on the \pt are reflected to the digital shadow automatically. Vice versa, the digital shadow does not change the state of the \pt.
\end{definition}
\end{tcolorbox}

\begin{figure*}[ht]
    \centering
    \includegraphics[width=0.9\textwidth]{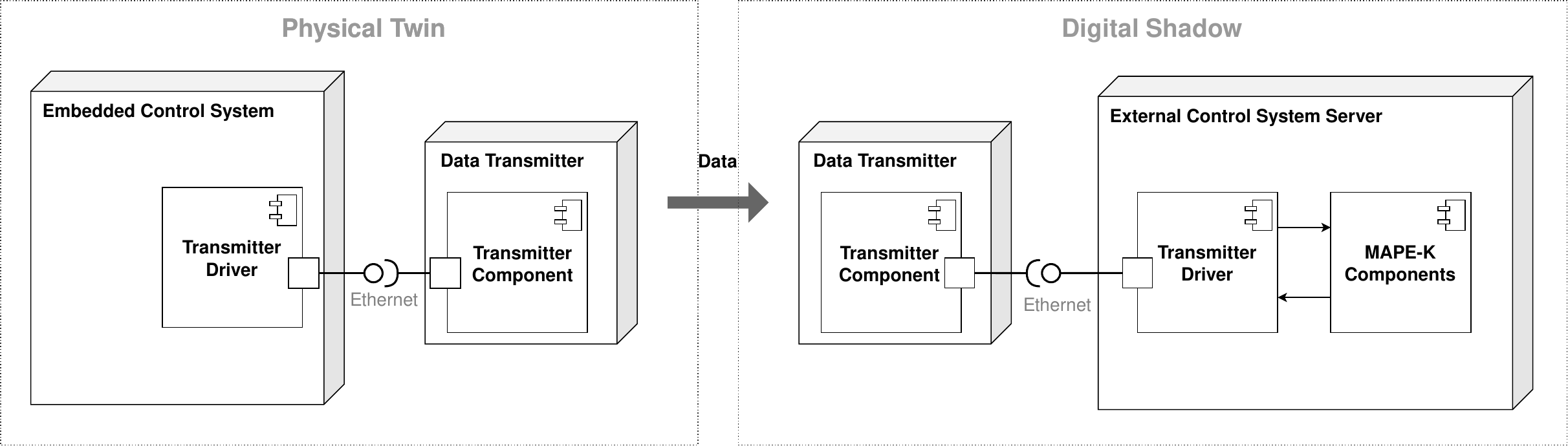}   
    \caption{The digital shadow is deployed separately from the \pt. The automated communication is unidirectional from the \pt to the digital shadow. Status changes and all other data is sent by the \pt and received by the digital shadow via transmitters. The digital shadow can reuse the transmitter driver from the \pt. The logic inside the digital shadow is based on the MAPE-K model.}
    \label{fig:systemds}
\end{figure*}

The configuration of the digital shadow for the \pt, as specified previously, is illustrated in \Cref{fig:systemds}. It is important to note that some parts of the \pt are not depicted in the figure. The digital shadow operates on a server that establishes a network connection to the \pt, either through a cable or wireless. In this example, assume a wireless connection between the \pt and its digital shadow. As the UML class diagram in \Cref{fig:umlallds} shows, many classes from the \pt can be reused. The transmitter uses the same device driver as the \pt, the event handlers are equal, and also the message types can be reused. Only the classes for the Monitor and Analyze stages of the MAPE-K model are new. A direct association between the two classes is not required, as they exchange data via an Observer pattern using the event handlers. Software package to enhance these two classes, are ignored in this example.

For data retrieval, the digital shadow employs a connected transmitter. To facilitate transmitter operation, the \pt's transmitter device driver can be repurposed. All data is then transmitted from the driver to the MAPE-K components. It is worth mentioning that MAPE-K is not an obligatory component of the digital shadow; it is used only for distinguishing representations between CPS, digital shadows, and a \dt.

Since machines controlled by external computers/servers already exist in the form of CPS, it is essential to clarify the distinction between a digital shadow and a CPS. As illustrated in \Cref{fig:mapek-ds}, the digital model holds the same level of importance as Knowledge. However, a CPS does not necessarily have to include a model of the connected machine, and even if it does, this model may not always be up-to-date. In contrast, for a digital shadow, this scenario is different. In the monitoring stage, all received data automatically updates the digital model.

\begin{figure}[ht]
    \centering
    \includegraphics[width=0.45\textwidth]{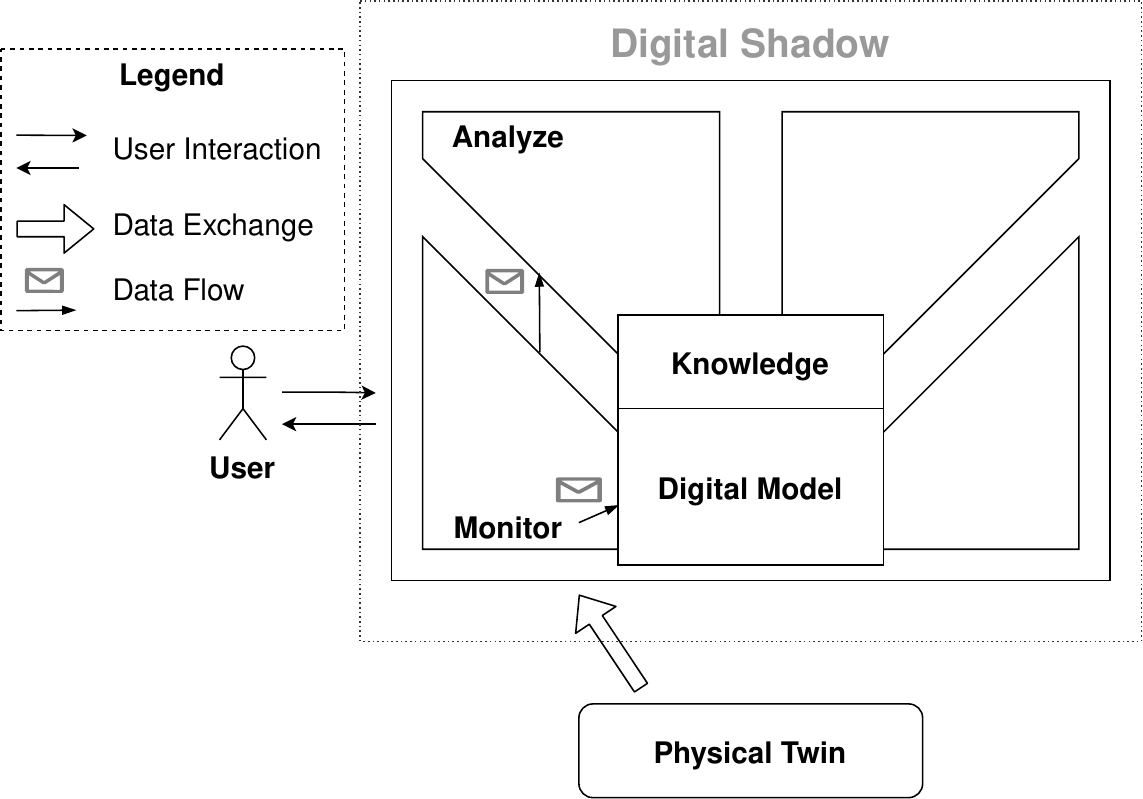}
    \caption{A digital shadow realized with the MAPE-K reference model. The Plan and Execution stages are not included, since there is also no data exchange from the Execution stage to the \pt.}
    \label{fig:mapek-ds}
\end{figure}

Another distinction is that a CPS can be used to directly operate the physical object. In contrast, a digital shadow's sole purpose is to monitor the \pt and provide data for analysis, enabling insight into the received data. Consequently, the Planning and Execution stages of the MAPE-K model are not inherent components of the digital shadow. While they can be incorporated, the automated change of state in the physical object is not a function of the digital shadow.

\subsubsection{Object-Z Formalization}
\begin{figure}[ht]
    \centering
    \includegraphics[width=0.3\textwidth]{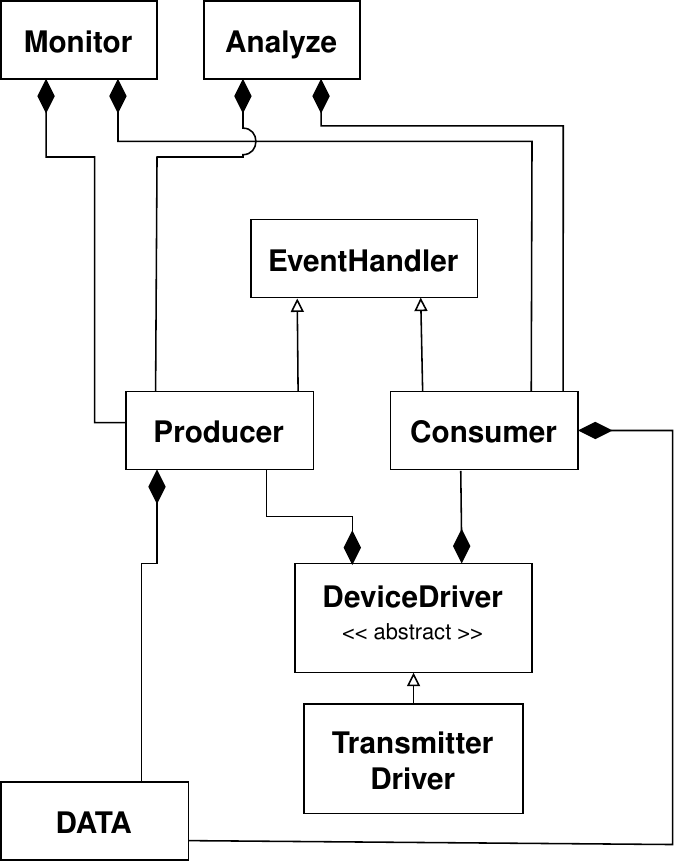}  
    \caption{Reduced UML class diagram of the digital shadow. The MAPE-K stages \textit{Monitor} and \textit{Analyze} are included, all other classes and relationships are identical to the UML class diagram of the \pt in \Cref{fig:umlall} on Page~\pageref{fig:umlall}.}
    \label{fig:umlallds}
\end{figure}

The UML class diagram in \Cref{fig:umlallds} is reduced to the two new classes for the Monitor and Analyze stages. All other classes and relationships are identical to the UML class diagram of the \pt in \Cref{fig:umlall} on Page~\pageref{fig:umlall}. A direct association between the classes is not required, as they exchange data via an Observer pattern using the event handlers. Software packages to enhance these two classes, are again ignored in this example.

A digital shadow specification with Object-Z can be done as follows. The \textit{Transmitter} and its operation are managed by the corresponding \textit{TransmitterDriver}, both of which can be reused from the Object-Z formalization provided for the \pt earlier. Additionally, all exchanged messages and the \textit{EventHandler} can also be reused. What remains to be specified is the Monitor and Analyze stage of the MAPE-K reference model.
The monitoring class \textit{Monitor} is a \textit{DTDigitalThreadProcess} and comprises two separate consumers: one for statuses and another for measurements:

\begin{class}{Monitor}
\project (\Init) \\ \zbreak
DTDigitalThreadProcess \\ \zbreak
\begin{state}
    statuses: Consumer \copyright\\
    measurements: Consumer \copyright\\
    emitter: Producer \copyright\\
    digitalModel: DigitalModel \\
\end{state}\\  \zbreak
\begin{init}
    statuses.\Init \land measurements.\Init
\end{init} \\ \zbreak
handleState \sdef \semi status : statuses.queue \\ \qquad \quad \spot statuses.Consume  \\ \qquad \quad \semi digitalModel.ProcessEvent \semi emitter.Emit \\ \zbreak
handleMeasurements \sdef \semi data: measurements.queue \\ \qquad \quad \spot measurements.Consume \semi emitter.Emit
\end{class}

Any status changes occurring in the \pt are emitted as STATUS events, while all measurements are emitted as \textit{MEASUREMENT} events. An emitter-producer is responsible for transmitting all consumed events to any registered listener. The most crucial component here is the \textit{digitalModel}, which is an object of the previously specified \textit{EventStateMachine}.

All status changes are handled by the \textit{handleState} function, which reads all \textit{STATUS} messages from the queue and forwards them to the digital model (state machine) for event processing. Subsequently, the result of the state machine's operation is emitted to all registered listeners. Since measurements do not impact the state machine's state, they are individually read from the queue via the \textit{handleMeasurements} function and immediately relayed to all registered listeners. One such listener could be a database (part of the Knowledge state) responsible for storing all data.

It is worth noting that the \textit{digitalModel} could also be a separate process that registers as a listener and consumes the \textit{STATUS} messages. In this example, the direct reference in the \textit{Monitor} class was used for better demonstration purposes.

The \textit{Analyze} stage is also a \textit{DTDigitalThreadProcess} and can be a (semi-)automated stage of the MAPE-K model in the context of the digital shadow. In this particular example, the \textit{Analyze} stage serves a singular purpose, which is to verify whether the received state from the \pt aligns with the state of the digital model or not. The outcomes of this comparison can then be emitted to all registered listeners. One potential listener could be a service responsible for notifying a user if any disparities in states are detected. Nonetheless, independent from the MAPE-K model, the analysis from the monitored events could also be done manually by a user, since no further stage is following:

\begin{class}{Analyze}
\project (\Init) \\
DTDigitalThreadProcess \\ \zbreak
\begin{state}
    consumer: Consumer \copyright\\
    emitter: Producer \copyright\\
    digitalModel: DigitalModel \\
\end{state}\\  \zbreak
\begin{init}
    consumer.\Init
\end{init}\\  \zbreak
\begin{op}{CheckStatuses}
    ptState?: STATUS \\
    equal!: \bool
\end{op} \\ \zbreak
compare \sdef \semi message : consumer.queue \\ \qquad \quad \spot consumer.Consume \semi CheckStatuses \\ \qquad \quad \semi emitter.Emit \\
\end{class}

With these processes, the \textit{DigitalShadow} schema can be defined. Since the MAPE-K example is only used for a better visualization of the concept, we use a more generic schema definition for the digital shadow:
\begin{class}{DigitalShadow}
    \begin{state}
        digitalModel: DigitalModel \copyright \\
        DThreadProcesses: \power DTDigitalThreadProcess \\
        DTtoPTConnection: DTtoPTConnection \copyright \\
    \end{state}
\end{class}
Please notice that no data is sent from the digital shadow to the \pt. The \textit{DTtoPTConnection} solely receives data from the \pt.

\subsection{The Digital Twin}\label{subsec:dt}
After defining and specifying the digital thread and digital shadow, the subsequent step is to comprehensively define the \dt. The \dt expands upon the digital shadow by enabling automatic synchronization of all alterations made to the digital model with the corresponding \pt. This means that any changes made to the \pt are mirrored in the \dt, and vice versa. Ultimately, the \dt evolves into a complete replica of the \pt. To formulate this definition, we draw upon the \dt definitions put forth by \textcite{dtdef-saracco} and \textcite{dtdef-trauer}:

\begin{tcolorbox}
\begin{definition}[Digital Twin]\label{def:dtdef-saracco}
A digital twin is a digital model of a real entity, the \pt. It is both a digital shadow reflecting the status/operation of its \pt, and a digital thread, recording the evolution of the \pt over time. The digital twin is connected to the \pt over the entire life cycle for automated bidirectional data exchange, i.e. changes made to the \dt lead to adapted behavior of the \pt and vice-versa.
\end{definition}
\end{tcolorbox}

\begin{figure}[ht]
    \centering
    \includegraphics[width=0.45\textwidth]{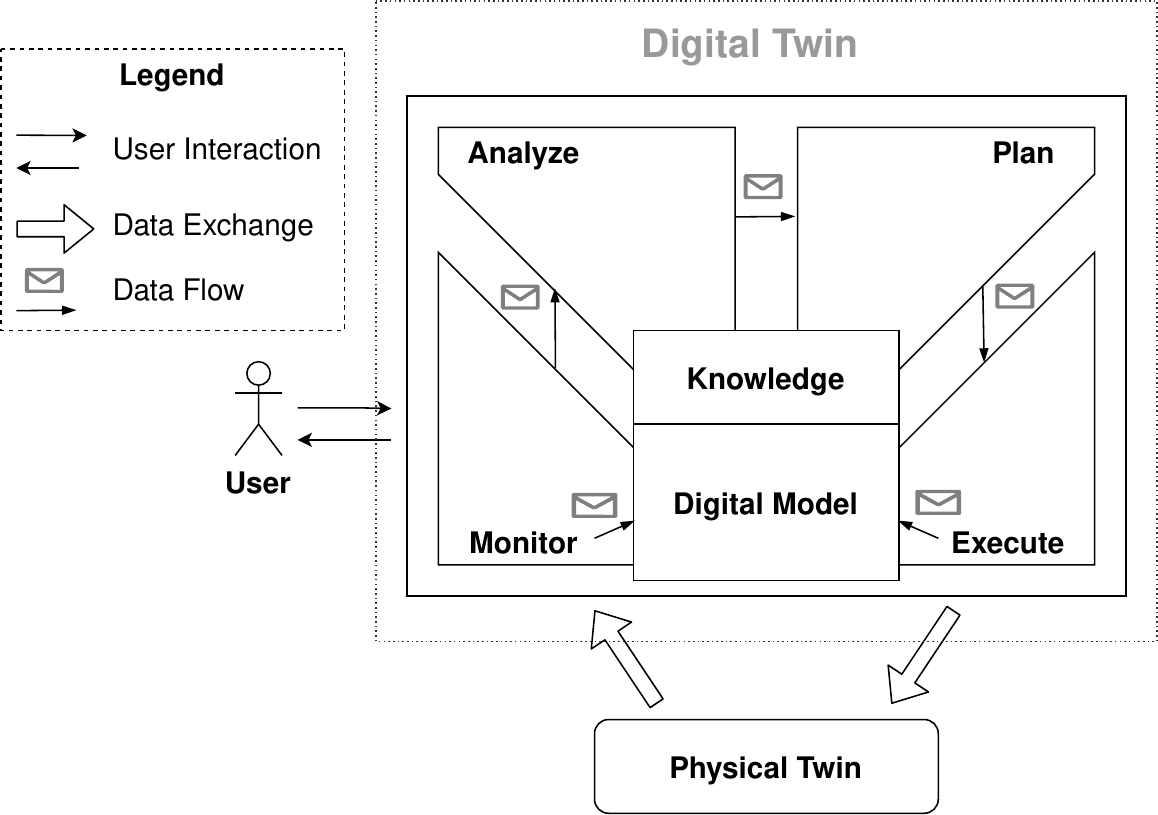}
    \caption{A digital twin realized with the MAPE-K reference model. The status change of the digital model and the corresponding data exchange from the Execution stage to the \pt is fully automated.}
    \label{fig:mapek-dt}
\end{figure}

Extending the system utilized in this example results in the addition of an extra communication channel from the \dt to the \pt, as illustrated in \Cref{fig:systemdt}. In the previously shown \Cref{fig:systemds}, the digital shadow only facilitates communication from the \pt to the digital shadow. Now, all modifications within the digital model are also transmitted from the \dt to the \pt.

\begin{figure*}[b]
\centering
\includegraphics[width=0.9\textwidth]{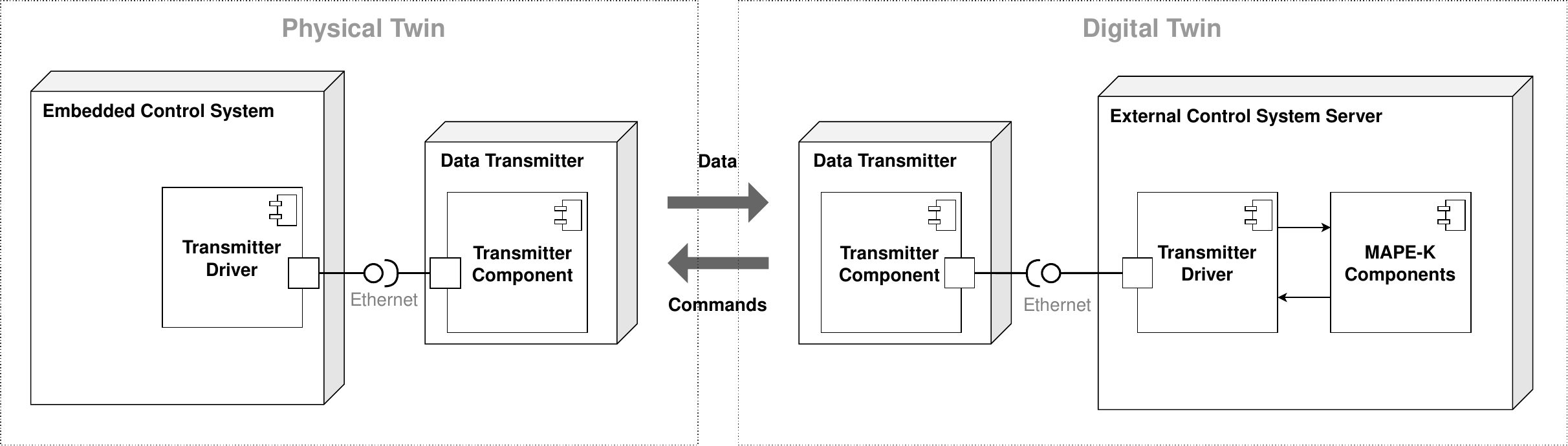} 
\caption{The digital twin extends the digital shadow in a way, that the communication between \pt and \dt is bidirectional. Additional to communication from the \pt to the \dt, all changes in the \dt are automatically sent to the \pt.}
\label{fig:systemdt}
\end{figure*}

Moreover, the MAPE-K model must be adapted to accommodate the \dt, as depicted in \Cref{fig:mapek-dt}. The Monitor and Analyze stages in this new model are identical to those in the digital shadow, as shown in \Cref{fig:mapek-dt}. The Plan stage takes the analysis results and formulates an execution scenario for the Execution stage if changes to the \pt are necessary. The key distinction from the original MAPE-K reference model lies in the \dt, where the Execution stage interacts with the digital model. Only if a positive result is returned, the command is sent to the \pt. Consequently, the digital model serves as the final control instance, and all incoming and outgoing changes are verified against the digital model.

\begin{figure}[h]
    \centering
      \includegraphics[width=0.35\textwidth]{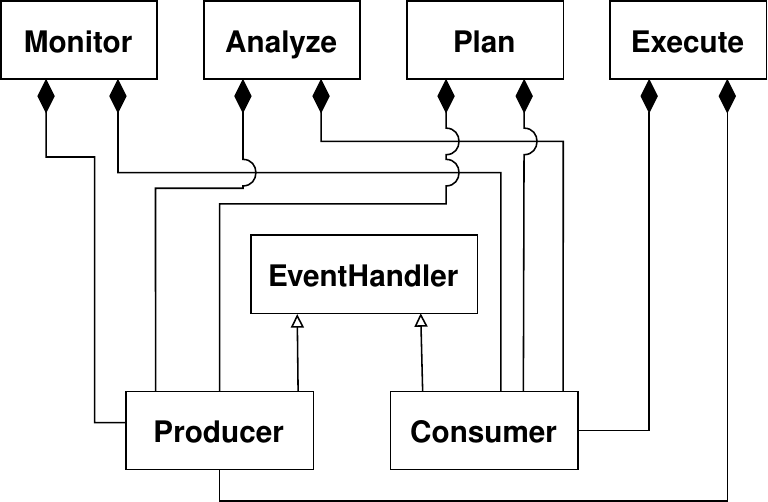} 
    \caption{UML class diagram of the digital twin, including only the MAPE-K relevant classes Monitor, Analyze, Plan, Execute, and the EventHandlers used for data exchange. All other classes are identical to the UML class diagram of the digital shadow in \Cref{fig:umlallds}.}
    \label{fig:umlalldt}
\end{figure}
\subsubsection{Object-Z Formalization}
The Object-Z formalization of the \dt can be built upon the digital shadow, incorporating two additional stages of MAPE-K as mentioned previously. First, the \textit{Plan} class is introduced: 

\begin{class}{Plan}
\project (\Init) \\
DTDigitalThreadProcess \\
\begin{state}
    consumer: Consumer \copyright\\
    emitter: Producer \copyright\\
    digitalModel: DigitalModel \\
\end{state}\\  \zbreak
\begin{init}
    consumer.\Init
\end{init}\\  \zbreak
\begin{op}{Planning}
    data?: \poly DATA \\
    result!: COMMAND
\end{op} \\ \zbreak
plan \sdef \semi message : consumer.queue \\ \qquad \quad \spot consumer.Consume \semi Planning \\ \qquad \quad \semi emitter.Emit \\
\end{class}

This class is also a \textit{DTDigitalThreadProcess} and includes a \textit{Consumer} component to receive data from the Analyze stage. All results generated during the planning stage are emitted via the \textit{Producer} \textit{emitter}. Similar to the other stages, the \textit{Plan} stage has direct access to the \textit{digitalModel}. However, in this example, no specific access details are provided.

The primary objective of this stage is to formulate a plan outlining which part of the \pt's software needs modification and how those modifications should be implemented. This task is executed through the \textit{plan} function. All incoming data is consumed and subsequently passed to the \textit{Planning} function. The resulting plan is then emitted to all registered listeners.

The last \textit{DTDigitalThreadProcess} is the \textit{Execute} class, which is kept straightforward as well. It receives all plans from the previous stage through the \textit{execute} function. The commands are validated against the \textit{digitalModel}, and the outcome is sent to the \pt. The \textit{transmitter} producer emits the command as an event to the \textit{TransmitterDriver}, which subsequently consumes this command and transmits it to the \pt:

\begin{class}{Execute}
\project (\Init) \\
DTDigitalThreadProcess \\ \zbreak
\begin{state}
    plans: Consumer \copyright\\
    transmitter: Producer \copyright\\
    digitalModel: DigitalModel \\
\end{state}\\  \zbreak
\begin{init}
    plans.\Init
\end{init}\\  \zbreak
\begin{op}{ChangeState}
    command?: COMMAND \\
    newState!: \poly STATE
\end{op} \\ \zbreak
execute \sdef \semi plan : plans.queue \spot plans.Consume \\ \qquad \quad \semi ChangeState \semi transmitter.Emit \\
\end{class}

Please note that the concrete implementation of the digital model in this context is not critical. The digital model could exist as a separate process that receives events through consumers and provides responses via producers. Alternatively, it could collect all events from the \textit{Execute} stage and independently transmit the results to the transmitter. There are numerous ways to realize this concept; however, the fundamental idea remains constant: changes to the digital model automatically trigger changes in the state of the \pt, without requiring any user intervention.

Similar to the digital shadow, we again define a generic schema \textit{DigitalTwin} without the MAPE-K processes:
\begin{class}{DigitalTwin}
    \begin{state}
        digitalModel: DigitalModel \copyright \\
        DThreadProcesses: \power DTDigitalThreadProcess \\
        DTtoPTConnection: DTtoPTConnection \copyright \\
    \end{state}
\end{class}
The schemes \textit{DigitalShadow} and \textit{DigitalTwin} look similar in this Object-Z formalization. The main difference is that the \dt can send state changes automatically to the \pt. 

\subsection{The Digital Twin Prototype}
Today's existing modeling and simulation tools can rapidly create a \dt of a single component or process, and \pubsub architectures 
allow all messages between processes to be captured and sent to a database or an IoT platform. However, complex \ifz applications 
require the integration of multiple sensors and actuators into a larger system, posing a challenge with no simple solution yet. 
The embedded community still uses various industrial interfaces and communication protocols such as ProfiBus, ProfiNet, ModBus, 
CANOpen, OPC-UA, or MQTT, to name a few. Some are proprietary, making integration difficult, for instance, ProfiBus and 
ProfiNet.

Robust software testing for communication protocols is challenging due to the difficulty of emulating or simulating them. Software 
engineers frequently use mock-up functions in unit tests to avoid the expensive networking exchange of data between processes, allowing them to obtain 
expected values. However, even robust unit testing with comprehensive edge case coverage is insufficient. Therefore, some approaches 
use simulation tools that replace the communication protocols between hardware components with software interfaces. For \ifz 
applications, both approaches are inadequate, as insufficient testing can jeopardize the safety of human operators. Despite this, 
simulation tools are crucial for the development of \ifz applications as a source of data for sensors and actuators. 

The software part of the connection can be formalized as shown in the \textit{Communication} schema. The physical part, however, where the data is sent between \textit{Device} 
and \textit{DeviceDriver} cannot be replaced in the same way. Hence, the approach still involves real hardware in the development 
loop. During development and testing, the \textit{Connection} object is the central piece. Without a counterpart, no command is 
executed, and no data is exchanged. Thus, engineers always require the hardware connected to the embedded software system they 
develop and test. Replacing the \textit{Connection} with a software mockup to circumvent \hil would result in a different 
\textit{Connection} object than used by the original \textit{SensorDriver}. Thus, the configuration during development would differ 
from the real counter part it is deployed on later. Furthermore, not all communication protocols used in industry are properly 
mockable. This can be demonstrated by the example of ModBus and OPC-UA applications on the OSI-Model shown in \Cref{fig:isoosi}. 
Unlike Ethernet-based communication protocols that implement and cover all layers of the OSI-Model, communication protocols based 
on serial connections, such as ModBus or CANOpen, are placed on the model's 7th layer, the \textit{Application Layer}. 
No additional host layers exist. Sending/receiving data is handled immediately by the \textit{Data Link} and \textit{Physical Layers}. 
This means that the physical hardware handles the necessary actions required for data exchange. Mocking these layers is difficult. 
On the other hand, communication protocols based on TCP, such as OPC-UA, can easily be mocked by opening a socket on the TCP layer 
and connecting another device to it. For serial protocols, this is not true. On connection, the driver tries to establish a 
connection to another device via RS232. As no device is connected, this would fail, and a connection error would be thrown.

Replacing the entire \pt during development and testing, which includes the hardware interfaces, leads to a fully virtual representation of the \pt and engineers do not necessarily need the hardware anymore for development. This is the main difference to the \dtp definitions by \textcite{dtdef-grieves, grieves2023}. We define the \dtp as follows:

\begin{tcolorbox}
\begin{definition}[Digital Twin Prototype]
A Digital Twin Prototype (DTP) is the software prototype of a \pt. The configurations are equal, yet the connected sensors/actuators are emulated. To simulate the behavior of the \pt, the emulators use existing recordings of sensors and actuators. For continuous integration testing, the DTP can be connected to its corresponding digital twin, without the availability of the \pt.
\end{definition}
\end{tcolorbox}

\begin{figure}[ht]
    \centering
    \includegraphics[width=.45\textwidth]{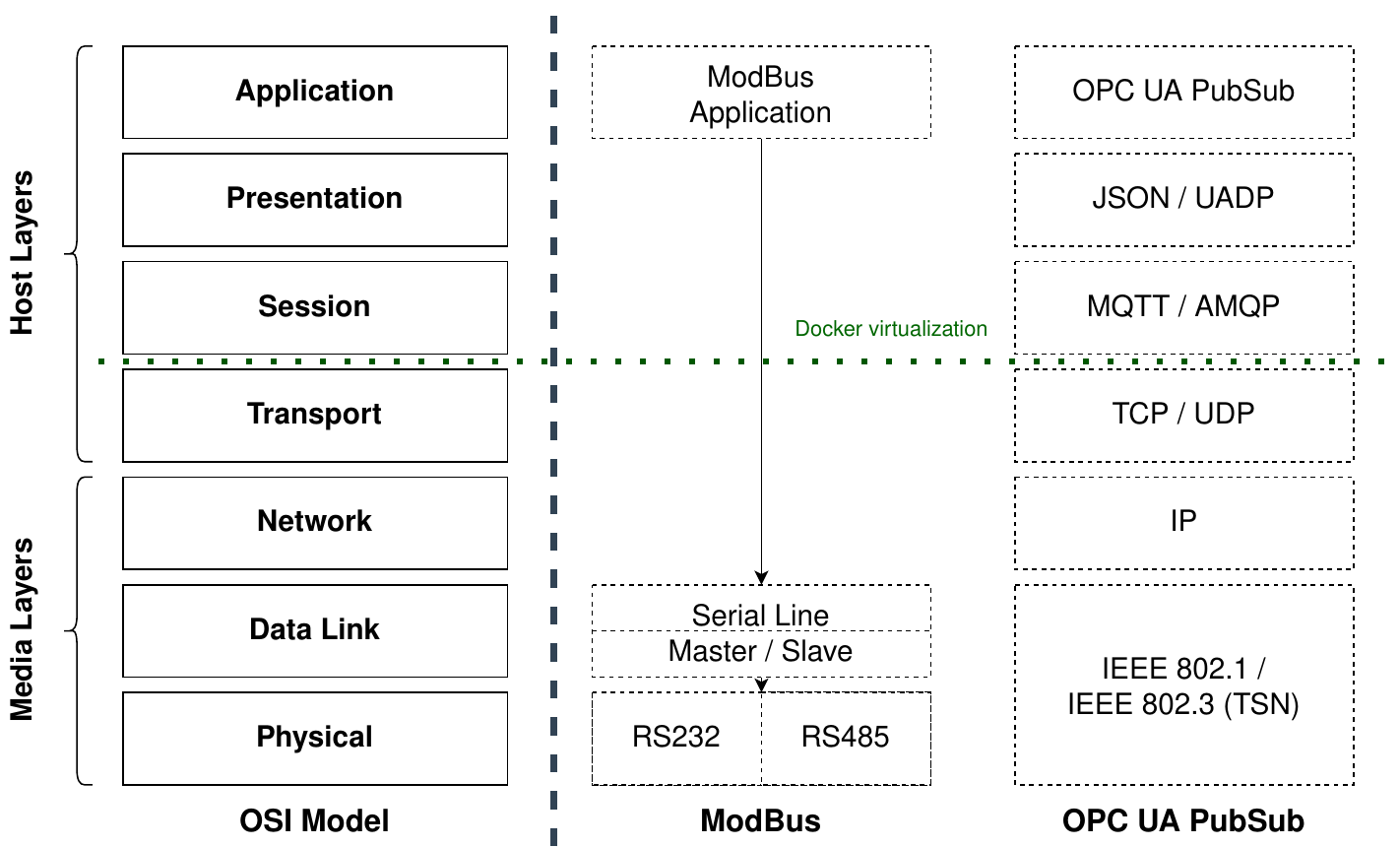}
    \caption{RS232 applications and communication can be visualized on the OSI layered model. 
    The application on Layer 7 is directly connected to the RS232 API and driver (Layer 2) that uses the physical connection 
    (Layer 1) to transmit data to other RS232 interfaces.}
    \label{fig:isoosi}
\end{figure}

\subsubsection{Object-Z Formalization}
To reduce the dependency of the embedded software system on the hardware during development and 
testing, communication protocols such as RS232 need to stay on the host layers of the OSI-Model without the need of changing the 
original connection properties of a device driver. This circumvents the layers that include the hardware. 
However, rerouting the connection disconnects the device and its driver. 
The rerouting only works if another process exists at the other end of the connection. So far, there is none. 
That is why not only the connection has to be emulated, but also the device. To begin, the emulated connection is defined first. 
The Object-Z formalization for \textit{EmulatedConnection} is as follows:

\begin{class}{EmulatedConnection}
\project (\Init, Read, Send, EmulateWrite, EmulateRead)\\
Connection\\ \zbreak
\begin{state}
    originalProtocol: \poly Connection
    \where
    originalProtocol \notin Connection
\end{state} \\ \break
\begin{init}
    type = TCP \\
\end{init} \\  \zbreak
\begin{op}{EmulateWrite}
data? : \poly DATA \\
forwardData!: \poly DATA
\where
forwardData! = data?
\end{op}\\  \zbreak
\begin{op}{EmulateRead}
data? : \poly DATA \\
forwardData!: \poly DATA
\where
forwardData! = data?
\end{op}
\end{class}

The \textit{EmulatedConnection} object inherits from the abstract \textit{Connection} class, and thus has all its properties and 
functions. This is shown on the OSI-Model in \Cref{fig:isoosi}. The safe way to stay in the host layers is to route all other 
communication protocols to \textit{TCP} and from there again back to the original protocol. 
Hence, the \textit{EmulatedConnection} does not replace the connection objects of \textit{Device} and \textit{DeviceDriver}. 
Instead, it is an independent additional connection that provides interfaces for a device emulator and a device driver to connect 
to with their original protocols. The \textit{EmulatedConnection} then uses TCP and forwards all incoming data via the function 
\textit{EmulateRead} and all outgoing data via the function \textit{EmulateWrite} between the emulated device and device driver.

How can this be realized without reconfiguring the device or device driver? Simply by using tools such as \textit{socat} (SOcket CAT) \cite{socat}. 
\textit{Socat} is a command-line utility that allows for bidirectional data transfer between two endpoints, typically over a 
network or through pipes. It is similar to the more well-known tool \textit{netcat}, but with 
support for multiple connection types and protocols (TCP, UDP, SSL, PTY, etc.). With two virtual serial ports (client and server) 
via \textit{socat} for the emulator and the device driver, a connection can be established without the need to change the 
configuration. In the background, \textit{socat} forwards the data between the ports via a TCP connection.

A device emulator for a sensor could be like the one shown in \Cref{fig:componentssensoremulator}. 
Similar to the real sensor, the \textit{SensorEmulator} inherits all properties and functions from the generic \textit{Device} class. 
There is only one difference; instead of executing a command and responding with the real result, the emulator uses virtual context 
for the response. Virtual context can be a list of previously recorded data from the real device or context provided by a simulation. 
In this example, we assume that the virtual context is previously recorded data with the real device.
\begin{figure}[ht]
\centering
        \subfloat[\label{subfig:sensorcomponent}]{%
      \makebox[2cm][c]{\includegraphics[width=0.15\textwidth]{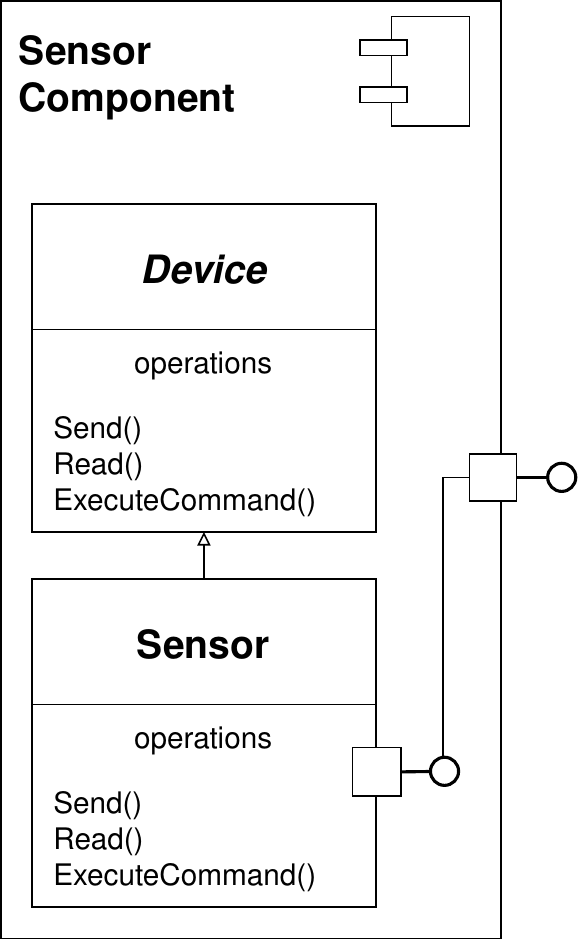}}
    }
    \hspace{0.5cm}
    \subfloat[\label{subfig:sensoremulatorcomponent}]{
      \makebox[2cm][c]{\includegraphics[width=0.15\textwidth]{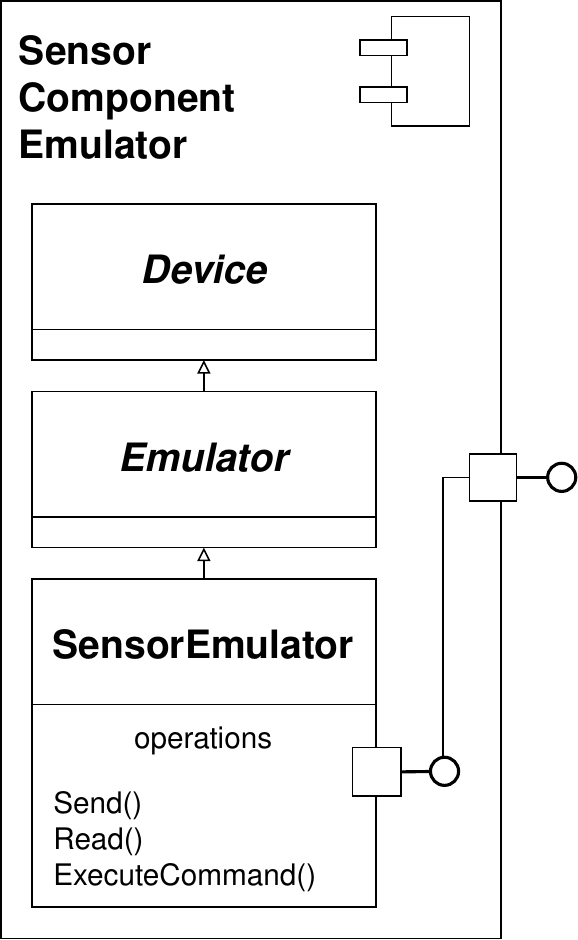}} 
    }
    \hspace{0.5cm}
    \subfloat[\label{subfig:sensordriver}]{
      \makebox[2cm][c]{\includegraphics[width=0.15\textwidth]{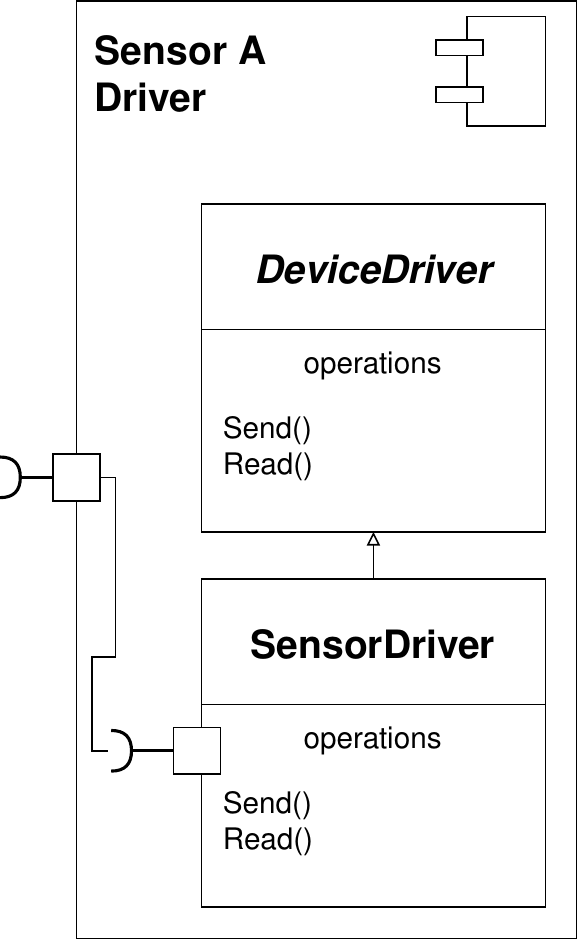}} 
    }
    \caption{UML component diagrams for sensor and emulator components. The real SensorComponent in (a) can be replaced by an EmulatedSensorComponent (b) and the SensorDriver (c) cannot distinguish whether it is connect to the real sensor in (a) or the emulated on in (b).}\label{fig:componentssensoremulator}
\end{figure}
Formalizing the emulated device and connection with Object-Z requires the definition of another data subtype first. 
Since the sensor responds to commands with a \textit{RESPONSE} type, a subtype of \textit{RESPONSE} named \textit{RECORDING} 
can be defined:
\begin{class}{RECORDING}
RESPONSE \\
\end{class}
The abstract class \textit{Emulator} inherits all properties and functions from the abstract class \textit{Device}, 
and \textit{SensorEmulator} inherits from \textit{Emulator}:
\begin{class}{Emulator}
Device\\
\end{class}

Although it may seem more obvious to inherit from \textit{Sensor}, 
the emulator cannot inherit its properties and functions from there. Most devices are a black box for the developer, and vendors 
only provide a technical manual and support to interact with the device. Thus, an emulator only mimics the behavior of the real 
counterpart and provides its API with corresponding return values. However, this is enough to replace the real device with the 
emulator for development and testing. A developer is mostly interested in the connection and data exchange part, not the internal 
behavior of a connected device. Due to abstraction reasons, the \textit{Sensor} object in this example was very simple. 
That is why the \textit{SensorEmulator} can also inherit all properties from \textit{Emulator} and change the \textit{ExecuteCommand} 
function to always return \textit{RESPONSE} objects from the \textit{virtualContext} set:

\begin{class}{SensorEmulator}
Emulator\\
    \begin{state}
    virtualContext: \power RECORDING
    \end{state} \\  \zbreak
\begin{op}{ExecuteCommand}
\Delta (virtualContext) \\
    command?: COMMAND \\
    result!: RECORDING
    \where
    command? \in commandList \\
    result! \in virtualContext  \\
    virtualContext' = virtualContext \setminus \{ result! \}
\end{op}\\ 
\end{class}
The \textit{SensorDriver} remains as it is and does not need any changes. The communication between an emulator and the 
\textit{SensorDriver} can be specified as follows using \textit{EmulatedCommunication}:
\begin{class}{EmulatedCommunication}
\begin{axdef}
    emulator: SensorEmulator \\
    driver: SensorDriver \\
    connection: EmulatedConnection
    \where
    \forall x: emulator.commandList \\ \qquad \quad \spot x \in driver.commandList \\
    \forall x: driver.commandList \\ \qquad \quad \spot x \in emulator.commandList
\end{axdef}\\  \zbreak
ToDrv \sdef emulator.Send \semi connection.EmulateWrite \\ \qquad \quad \parallel connection.EmulateRead \semi driver.Read \\
ToDev \sdef driver.Send \semi connection.EmulateWrite \\ \qquad \quad \parallel connection.EmulateRead \semi emulator.Read \\
\end{class}
The \textit{EmulatedCommunication} object now includes an additional \textit{Connection} object in the form of 
\textit{EmulatedConnection}. The communication from the emulator to the device driver, labeled as \textit{ToDrv} is now a 
composition of the connections from the device to the \textit{EmulatedConnection}. From there, the data is sent to the device driver, 
where the \textit{EmulatedConnection} receives it and forwards it to the connection defined by the device driver. 
The \textit{EmulatedConnection} is not part of either the device/emulator or the device driver. 
Therefore, in this example, the \textit{SensorDriver} cannot differentiate between whether it is connected to a real device or an emulator, 
which is the goal of our approach.

\subsection{Summary of the Digital Twin Concept}
\begin{figure*}[b]
    \centering
    \includegraphics[width=.8\textwidth]{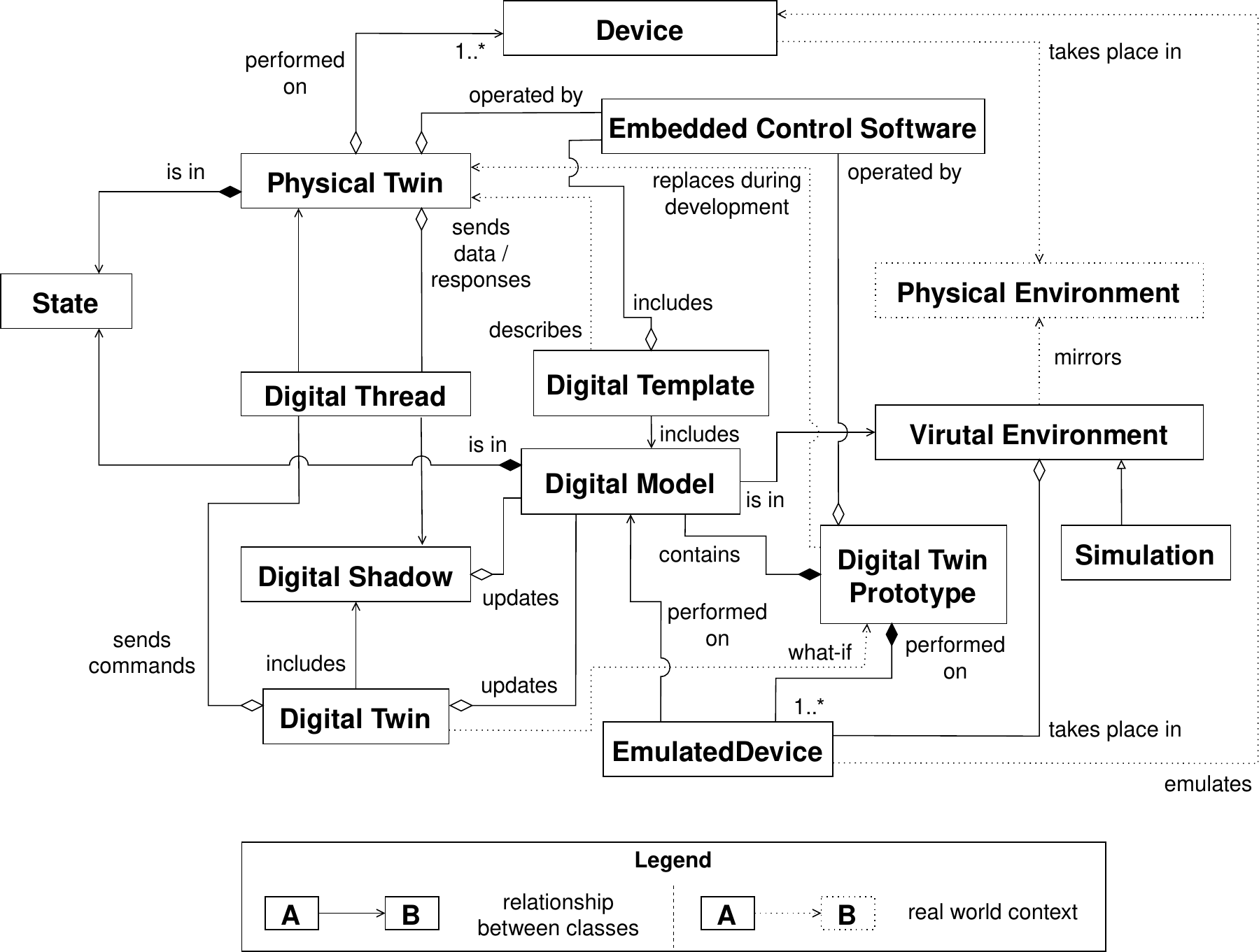}
    \caption{Relationships between \pt, digital model, digital template, digital shadow, \dt, and \dtp.}
    \label{fig:relationships}
\end{figure*}

The relationships between the different concepts are illustrated in the UML diagram in \Cref{fig:relationships}. We extended the semi-formal approaches by \textcite{Yue2021} and \textcite{Becker2021} for the digital twin the digital shadow. A \textit{Physical Twin} performs actions using real \textit{Devices} in a \textit{Physical Environment}. The \textit{Physical Environment} is not a real class, but the real world context in which the \textit{Device} operates. Changing behaviors lead to changes in the current state of the physical twin. Hence, the physical twin updates its state and sends the change of state via the \textit{Digital Thread}, which was named \textit{Twinning} in \textcite{Yue2021}, to the \textit{Digital Shadow}. Different to the formalization by \textcite{Yue2021}, the \pt is not directly connected to the \dt, but via the \textit{Digital Shadow}, which is included by the \dt. In our Object-Z formalization of the digital shadow and \dt, we illustrated the difference utilizing the MAPE-K model and showed that the digital shadow does not send any data to the \pt. All state changes are received by the digital shadow, which then changes the \textit{Digital Model}. Only the \textit{Digital Twin} updates state changes similar to the change of state of the \textit{Physical Twin}. Instead of physical processes, the \dt uses the \textit{Digital Model}, which operates in a \textit{Virtual Environment}, to change the \pts state. During the development phase, the \textit{Digital Twin Prototype} can replace the \pt. A \dtp executes commands on \textit{Emulated Hardware} in a \textit{Virtual Environment}. The \textit{Virtual Environment} should mirror the real world, which can be realized via a \textit{Simulation}. To describe and construct the \textit{Physical Twin} its \textit{Digital Template} can be used, since it includes the \textit{Digital Model} and the \textit{Embedded Control Software}.

The special feature of the \dtp is that it is operated by the same \textit{Embedded Control System} as the \pt. This software does not even recognize, whether physical hardware or emulated hardware is used. Notice that the \textit{Digital Model} used by the \dtp is a different instance than the \textit{Digital Model} updated by the \textit{Digital Shadow}. Advanced \textit{Digital TWins} can use the \textit{Digital Twin Prototype} to evaluate ``what-if'' questions in more realistic scenarios that include the full software stack.

\section{Application of this Concept}
In the following, two projects are presented, where the previous definitions and methods were already applied in real life contexts.

\subsection{Field Experiment with Underwater Ocean Observation Systems}\label{sec:arches}
The \dtp approach was developed for a network of ocean observation systems and tested during the research cruise AL547 with RV ALKOR (October 20-31, 2020) of the Helmholtz Future Project ARCHES (Autonomous Robotic Networks to Help Modern Societies) \cite{demomission}. In ARCHES, with a consortium of partners from AWI (Alfred-Wegener-Institute Helmholtz Centre for Polar and Marine Research), DLR (German Aerospace Center), KIT (Karlsruhe Institute of Technology), and the GEOMAR (Helmholtz Centre for Ocean Research Kiel), several \dtps for ocean observation systems were developed. The major aim of this project was to implement robotic sensing networks, which are able to autonomously respond to changes in the environment by adopting its measurement strategy, in both space and in the deep sea. A field report on employing \dtps in this context is published by \textcite{demomission}.

Five \dtps of ocean observation systems constructed at AWI and GEOMAR were developed. They vary in construction, payload, and configuration. The distance between AWI and GEOMAR are a few hundred kilometers. Hence, the \dtps were used to develop the software, without a permanent connection to the physical ocean observation systems. The microservices were implemented with ROS and encapsulated in Docker. How the different \dtps of the ocean observation systems were developed, was describe by \textcite{MFI2020}. A special feature in this project was that the \dtps were used as \dts of the \pts underwater. The fully virtualized embedded software systems showed the state of the \pts. This way, no extra software to run a \dt was required.

Furthermore, with \dtps it was possible to develop and test scenarios before the mission took place. Automated testing is implemented through CI/CD in Gitlab. During the mission, all exchanged messages on the \dt and \dt were recorded and can now be used to increase the quality of the CI/CD pipelines.

\subsection{Case Study with Smart Farming Applications}\label{sec:silolytics}
As the digitalization of agricultural processes promotes the use of \dts for various use cases \cite{nasirahmadi2022review}, we also report on a case study that experimented with the \dtp approach for a smart farming application.

The smart farming project SilageControl with a consortium of the Silolytics GmbH (project lead), Blunk GmbH, and Kiel University used \dts to adopt the \dtp approach for development and maintenance. The major goal of SilageControl is to improve the process of silage making, i.e. the fermentation of grass or corn in silage heaps. In order to avoid mold formation, the harvested crop is compacted by heavyweight tractors.
As displayed in \Cref{fig:silagecontrolfg}, these tractors are equipped with a sensor bar, which includes GPS sensors, an inertial measurement unit (IMU), and a LiDAR. In combination, the sensors enable the continuous and accurate representation of the tractor's position / orientation and the shape and volume of the silage heap.

\begin{figure}[h]
\centering
    \subfloat[Sensor bar in lab environment\label{subfig:scbar}]{%
      \includegraphics[width=0.4\textwidth]{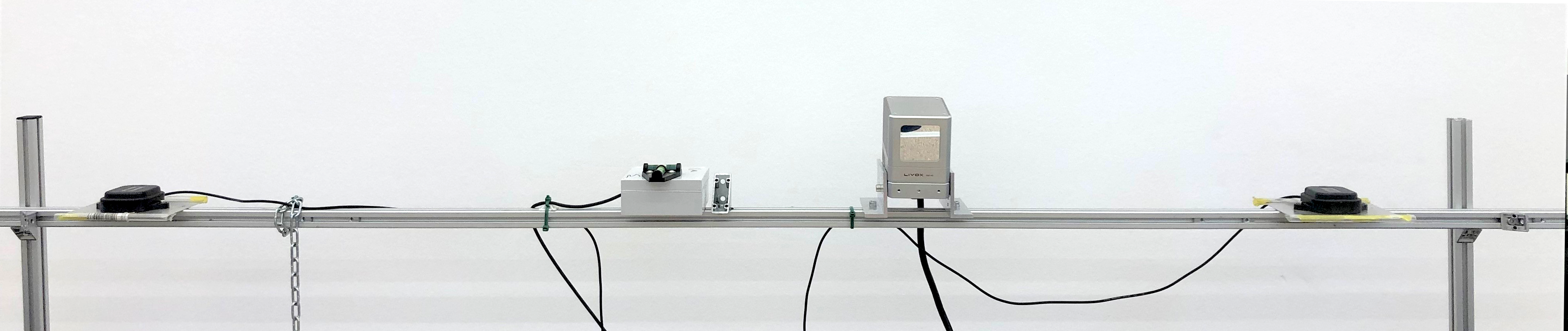} 
    }
    \\
    \subfloat[Sensor bar mounted on a tractor \label{subfig:sctractor}]{%
      \includegraphics[width=0.4\textwidth]{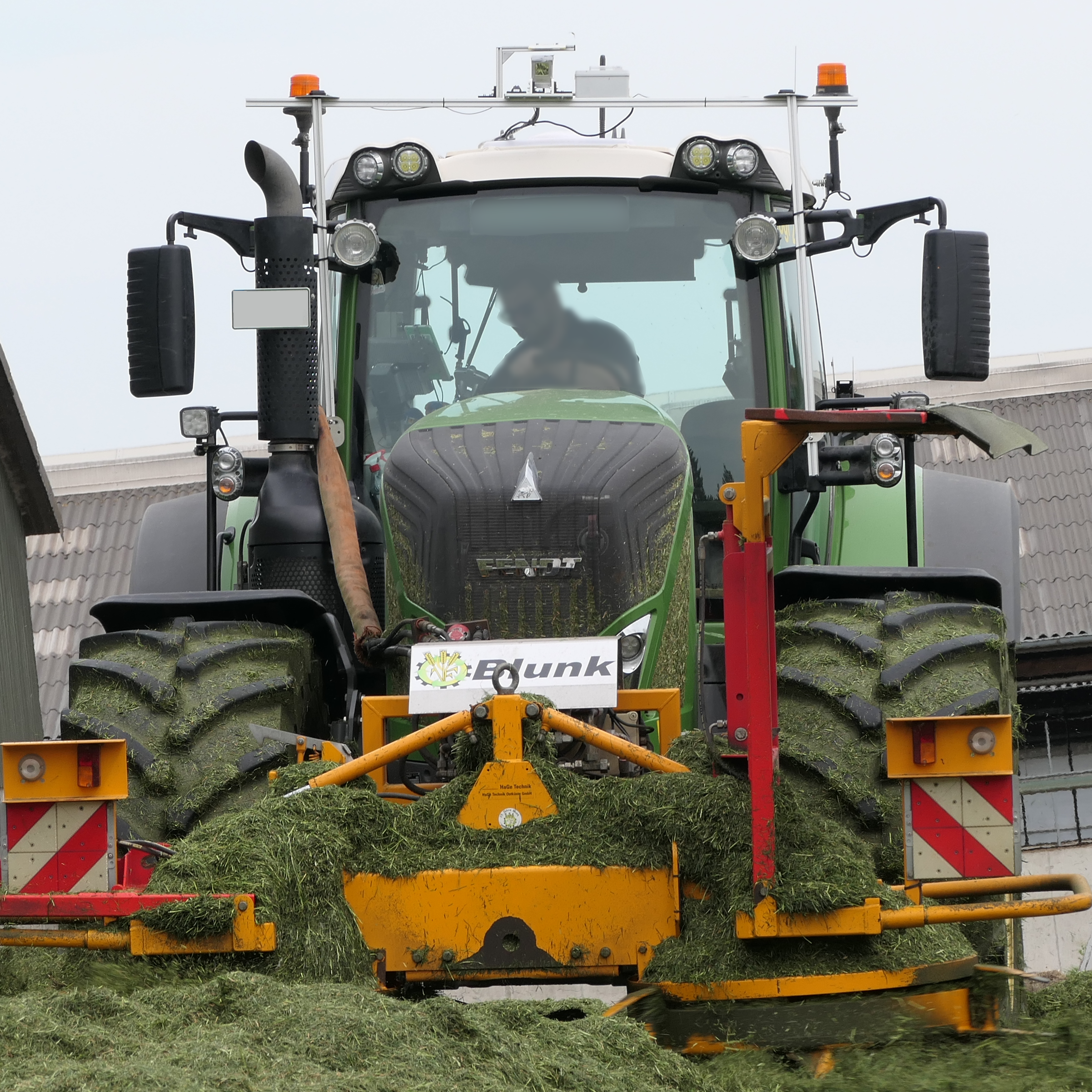} 
    }
    \caption{Sensor bar which monitors the process of silage making.}\label{fig:silagecontrolfg}
\end{figure}

Since silage making is season dependent, the \dtp approach is used to improve the sensor platform independent from the current season. The first field experiments were conducted from May to October 2022. During this period, sensor data was be recorded to further improve the accuracy of physical models and create scenarios for automated testing of future features. Thereby, data gathered by the \dt improves the \dt/\dtp and vice versa. A case study with more details about this project was published by \textcite{silagecontrol}.

\section{Conclusion and Future Work}
Digital twins find applications across all layers in Industry 4.0 scenarios \cite{MFI2020}. However, there exists confusion in the definitions of digital models, digital shadows, digital twins, and digital twin prototypes. While many studies attempt to list and categorize these differences, a formal description has been lacking. Therefore, in our Digital Twin concept, we formally specified the various components, ranging from the physical twin to the digital twin, culminating in a fully virtualized digital twin prototype capable of substituting the physical twin during development. To underscore the distinctions among these different facets of the digital twin from a software engineering standpoint, we provide an Object-Z formalization for each component.

We extended the \dt concept by the \textit{Digital Template}. A digital template describes the \pt and is used to build it. It includes the \pt's \textit{Digital Model},  describing documents, and the \textit{Embedded Control Software}.

We have provided real-world application examples to illustrate the practical context. A proof of concept for the formal specifications was demonstrated in a demonstration mission showcasing the viability of digital twins in ocean observation systems \cite{demomission}. Moreover, we offered insight into how this approach could be employed in the SilageControl smart farming project, which aims to enhance the silage-making process through the development of a sensing platform \cite{silagecontrol}. 

The usage of \dtps transforms the way how embedded software systems are developed. By starting with the emulation of hardware sensor by sensor, actuator by actuator, and communication protocol by communication protocol, the development of embedded software systems becomes an iterative process. Furthermore, the integration of a fully operational \dtp heralds a shift towards collaborative efforts between engineers and domain experts, regardless of their physical location or connection to the hardware.

Besides reducing the time that is needed for testing by switching from HIL to SIL testing with \dtps, this approach also avoids expenses for redundant hardware and paves the way for more efficient development workflows that are otherwise difficult to implement for embedded software systems. Digital twins become a key enabler for fully automated integration testing of embedded software systems in CI/CD pipelines. While building, testing, and releasing of software is possible for embedded software just like in other fields of software engineering, integration testing with hardware interaction is expensive, due to the HIL testing, and is often done manually. Thus, the integration tests are a bottleneck in the verification and validation activities and, hence, the release of new software.
Anyway, with proper integration testing, developers increase the robustness of the embedded software systems. This may even embrace Industrial DevOps methods in the embedded field \cite{ICSA2019}.

In summary, digital twins have the potential to enhance the quality of embedded software systems, concurrently reducing costs and accelerating development speed. These benefits align with the challenges cited by both Ebert~\cite{SIsoftwarequality} and Ozkaya~\cite{50yearsse}], who identified the challenges to achieve quality while managing costs and efficiency.

Nevertheless, the \dt community still has a lot of home work to do. The lack of a consensual definition of \dts leads to a lot of room for interpretation what a \dt is. Instead of introducing abstract approaches that are described using an attached case study, researchers should focus more on formal approaches to demonstrate and distinguish different approaches. This still may leads to many different \dt definitions, but at least the community is able to consolidate similar approaches and has a starting point to discuss differences, flaws, or benefits of different approaches. With the introduction of virtualization tools such as Docker and open platforms such as GitHub, the distribution of code and tools to replicate results of a research study or experiment with an approach became easy and has no costs attached.

The validation of research results and the reproducibility of experiments are integral aspects of good scientific practice \cite{empiricalstandard}. However, replicating the conducted field experiments from our ARCHES demonstration mission or the SilageControl case study using similar hardware can be quite expensive. To facilitate independent replication of the \dtp approach by engineers and other researchers, we have developed a \dtp using cost-effective hardware, specifically a PiCar-X by SunFounder \cite{SunFounder}. This digital twin prototype is based on the ARCHES Digital Twin Framework \cite{ADTF} and is publicly available on GitHub \cite{abarbiegithub}. More comprehensive details about the PiCar-X digital twin prototype will be presented in a separate publication.

\printbibliography

\end{document}